\documentclass{emulateapj}
\usepackage{graphicx}
\usepackage{apjfonts}
\usepackage{color}
\usepackage{bm}
\slugcomment{\scriptsize{Submitted to ApJ, Preprint typeset using \LaTeX style}} 
\shorttitle{Differential Rotation in Magnetized and Non-magnetized Stars}
\shortauthors{Mabuchi, Masada and Kageyama}
\begin{document}
\title{Differential Rotation in Magnetized and Non-magnetized Stars}
\author{Jun Mabuchi\altaffilmark{1}, Youhei Masada\altaffilmark{1} and Akira Kageyama \altaffilmark{1}} 
\altaffiltext{1}{Department of Computational Science, Graduate School of System Informatics, Kobe University; 
Kobe 657-8501, contact: ymasada@harbor.kobe-u.ac.jp}
\begin{abstract}
Effects of magnetic field on stellar differential rotation are studied by comparing magnetohydrodynamic (MHD) models and their hydrodynamic 
(HD) counterparts in the broad range of rotation rate and in varying initial rotation profile. 
Fully-compressible MHD simulations of rotating penetrative convection are performed in a full-spherical shell geometry. 
Critical conditions for the transition of the differential rotation between faster equator (solar-type) and slower equator (anti-solar type) are explored 
with focusing on the ``Rossby number (${\rm Ro}$)" and the ``convective Rossby number (${\rm Ro}_{\rm conv}$)". It is confirmed that the transition 
is more gradual and the critical value for it is higher in the MHD model than the HD model in the view of the ${\rm Ro}_{\rm conv}$-dependence. 
The rotation profile shows, as observed in earlier studies, the bistability near the transition in the HD model, while it disappears when allowing the 
growth of magnetic fields except for the model with taking anti-solar type solution as the initial condition. We find that the transition occurs at 
${\rm Ro} \simeq 1$ both in the MHD and HD models independently of the hysteresis. Not only the critical value, the sharpness of the transition 
is also similar between the two models in the view of the ${\rm Ro}$-dependence. The influences of the dynamo-generated magnetic field and/or the 
hysteresis on convective motion are reflected in the ${\rm Ro}$.  This would be the reason why the transition is unified in the view of the 
${\rm Ro}$-dependence. We finally 
discuss the ${\rm Ro}$-dependence of magnetic dynamo activities with emphasis on its possible relation to the kinetic helicity profile.  
\end{abstract}
\section{Introduction}
The differential rotation (DR) is believed to be a key ingredient in organizing large-scale magnetic fields in the solar interior. Most of  
standard solar dynamo scenarios rely heavily on, so-called $\Omega$-effect as the amplification process of magnetic fields to 
reproduce observed solar magnetic activity with cyclic polarity reversals and butterfly-shaped spatiotemporal migrations 
\citep[see, e.g.,][for reviews]{charbonneau05,charbonneau10}. However, we have not yet arrived at a full understanding of 
the physical mechanism for maintaining the solar differential rotation. See, e.g., \citet{miesch05} for a review. 

The internal rotation profile of the Sun has been uncovered by the global helioseismology \citep[e.g.,][]{kosovichev+97,thompson+03,howe09}. 
A striking feature of the solar rotation profile is its equatorial acceleration: The plasma near the equator rotates faster than other parts in the 
convention zone. While the bulk of the convection zone is characterized by conical iso-rotation contours, the strong radial shear is 
concentrated in the near-surface layer and the tachocline underlying the convection zone \citep[e.g.,][]{schou+98,charbonneau+99,basu+01}. 
The equatorward angular momentum transport due to the turbulent Reynolds stress induced by the rotating stratified convection is mainly 
responsible for the solar-type DR with equatorial acceleration \citep[e.g.,][]{krause+74,rudiger89,kitchatinov+95,miesch05}. 

The possibility of the opposite profile, or anti-solar type DR, in which the equatorial region is rotating slower, has been suggested numerically 
since the early 3D hydrodynamic simulation of rotating spherical-shell convection by \citet{gilman77} over a broad range of parameters and 
various simulation setups \citep[see][]{glatzmaier+82,brun+02,aurnou+07,steffen+07,chan10,kapyla+11,gastine+13}. However, although the 
solar-type DR has been frequently observed in various types and ages of the stars, the anti-solar type DR has only been reported for a few 
of K giant stars observed with the Doppler imaging technique \citep{strassmeier+03,weber+05,kovari+07, kovari+14}. It is still unclear how 
common the anti-solar type DR is in the solar-like main-sequence dwarfs and what really separates the two rotation regimes. 

Bearing in mind these situations, the transition between the solar and anti-solar type rotation profiles has been extensively 
studied by the simulation of spherical shell convection in recent years. \citet{kapyla+11} found in fully-compressible convection simulation 
that the transition is characterized by the Coriolis number (inverse of the Rossby number which is defined in \S2) and is only weakly dependent 
on the density stratification. \citet{guerrero+13} explored the physics of large-scale flows in solar-like stars by using an anelastic large-eddy 
simulation with stratification resembling the solar interior. They confirmed that the two regimes of the DR exist even in the solar-like 
strongly stratified internal structure. The critical Rossby number they obtained was consistent with that in \citet{kapyla+11}. 

The systematic parameter study of the transition between the solar and anti-solar type rotation profiles was conducted by \citet{gastine+14} 
with rotating spherical shell simulations of anelastic and Boussinesq convections \citep[see also][]{gastine+13}. They found, by combining their massive 
simulation results with the models of the different research groups, that the transition is controlled by the ``convective Rossby number", 
almost independently of the model setup, such as thickness of the convective envelope and density stratification. In addition, they found that 
two kinds of DR profiles are two possible bistable states around the transition, suggesting the hysteresis of the stellar rotation profile. The 
bistability of the rotation profile, discovered by \citet{gastine+14}, was confirmed independently by \citet{kapyla+14} in hydrodynamic simulations 
of the compressible convection. However, the influence of magnetic field on the rotation profile and its transition is still controversial despite 
the magnetic field is an inevitable outcome of electrically-conducting fluid motions \citep[e.g.,][]{moffatt78,krause+80}. 

In \citet{masada+13} (hereafter MYK13), we suggested, for the first time, that the formation of the solar-type DR is associated with the 
development of the magnetic field : It was found that the anti-solar type DR profile at the early dynamo kinematic stage transits to 
the solar-type profile after the dynamo-generated magnetic field beginning to affect the convective motion (see \S4.1 in MYK13). 
Such a magnetic influence on the rotation profile was confirmed by independent groups recently. \citet{fan+14} performed anelastic convective 
dynamo simulation of the model with the realistic solar internal structure, solar rotation rate and solar luminosity. They showed that the 
dynamo-generated magnetic field plays a crucial role in attaining the solar rotation profile in the actual solar parameter regime, i.e., the 
rotation profile becomes the anti-solar type without the magnetic field. They also reported that the bistability of the 
rotation profile, discovered in earlier studies, disappears when allowing the evolution of the magnetic field. The facilitation of the solar-type DR and the disappearance of 
the bistability of the rotation profile due to the magnetic field were confirmed by \citet{karak+14}, in which they studied the effect of the magnetic 
field on the rotation profiles around the transition by fully-compressible convective dynamo simulations. 

In this paper, we systematically study the effect of the magnetic field on rotating spherical shell convection and resultant mean DR 
by comparing the MHD models with their hydrodynamics counterparts (HD models) in the broad range of the initial rotation rate. The bistable 
nature of the rotation profile is revisited under the influence of the magnetic field. A self-consistent, fully-compressible Yin-Yang MHD code 
which was developed in MYK13 and a stellar model consisting of the convection zone and the radiative zone are used for the simulation. 
The primary objective of this paper is to explore a key parameter which controls the transition between the anti-solar type and solar-type 
DR profiles and its dependence on the magnetic field and the initial rotation profile (\S3.1--3.3). We cast a spotlight on two diagnostic 
parameters, i.e., the ``Rossby number" and the ``convective Rossby number" (see \S2 for the definitions). The magnetic dynamo activity in the MHD 
 models with the different rotation rates is also studied in \S3.4. After discussing the relation between the magnetic dynamo activity and the 
 convection properties in \S4, we summarize our findings in \S5. 
\section{Simulation Setup}
\begin{table*}[tbp]	
\begin{center}
\caption{Summary of the simulation runs. The labels of MHD models contain the letter M. HD models 
have a label beginning with H. The second letter denotes the initial rotation profile. The labels of basic runs with the initial rigid rotation contains 
the letter R. Models started with a solar-type or an anti-solar type DR have a letter S or A. The number indicates the given rotation rate 
$\Omega_0$ with the decimal point removed. The mean quantities $v_{\rm rms}$, $B_{\rm eq}$, ${\rm Ro}$, $\tau_c$, 
$\epsilon_{\rm M}$, and $\epsilon_{\rm K}$ are evaluated at the saturated state in the convective envelope. The super-criticality level $\delta $ is defined by 
$(Ra - Ra_c)/Ra_c$, where $Ra_c \simeq 9.83E^{-1.16}$ with the Ekman number of $E \equiv \nu/(\Omega_0d_0^2)$. Here we adopt, 
as the critical Rayleigh number, the fitting formula derived by \citet{al-shamali+04}, in which they obtained it from the simulation of the 
spherical shell convection with varying the shell thickness and the Ekman number. 
$\alpha_e = \bar{v}_\phi (r = R,\theta = \pi/2)/(\Omega_0 R) $ is the DR parameter. 
In the final column, the resultant DR profiles, 
S (solar-type) or AS (anti-solar type), are summarized. } 
\begin{tabular}{c c c c c c c c c c c c } \hline\hline
  & $\Omega_0 $ & $\delta$ & $v_{\rm rms} $ & $B_{\rm eq}$ & $\tau_c$ & $\epsilon_{\rm K}$ & $\epsilon_{\rm M}$ & ${\rm Ro}$ & ${\rm Ro}_{\rm conv}$  & $\alpha_e$ &  type   \\ \hline\hline
MR00625 &0.0625 &$ 155 $  & $0.037 $ & $ 0.028 $ & $8.0$ & $1.3\times10^{-3}$ & $2.7\times10^{-4}$ & $6.272$ & $3.737$ & $-0.351$ & AS \\ 
MR0125 &0.125 &$ 69 $  & $0.040 $ & $ 0.028 $ & 7.6 & $1.6\times10^{-3}$ & $4.0\times10^{-4}$ & $3.310$ & $1.868$ & $-0.386$ & AS \\ 
MR025 &0.25 &$ 30 $  & $0.029 $ & $ 0.021 $ & 10.2 & $8.4\times10^{-4}$ & $8.8\times10^{-4}$ & $1.230$ & $0.934$ & $-0.094$ & AS \\ 
MR030 &0.30 &$ 24 $  & $0.028 $ & $ 0.021 $ & 10.5 & $8.3\times10^{-4}$ & $7.4\times10^{-4}$ & $0.995$ & $0.778$ & $-0.069$ & AS \\ 
MR035 &0.35 &$ 20 $  & $0.030 $ & $ 0.022 $ & 10.0 & $8.6\times10^{-4}$ & $2.6\times10^{-4}$ & $0.898$ & $0.667$ & $0.045$ & S \\ 
MR040 &0.40 &$ 17 $  & $0.029 $ & $ 0.021 $ & 10.3 & $8.3\times10^{-4}$ & $2.4\times10^{-4}$ & $0.763$ & $0.584$ & $0.086$ & S \\ 
MR050 &0.50 &$ 13 $  & $0.028 $ & $ 0.020 $ & 10.8 & $8.0\times10^{-4}$ & $2.5\times10^{-4}$ & $0.580$ & $0.467$ & $0.088$ & S \\ 
MR060 &0.60 &$ 10 $  & $0.027 $ & $ 0.020 $ & 11.3 & $7.4\times10^{-4}$ & $2.7\times10^{-4}$ & $0.463$ & $0.389$ & $0.078$ & S \\ 
MR150 &1.50 &$ 3 $  & $ 0.018$ & $ 0.012 $ & 17.0 & $2.8\times10^{-4}$ & $2.9\times10^{-4}$ & $0.123$ & $0.156$ & $0.011$ & S \\ 
MR300 &3.00 &$ 0.75 $  & $0.010 $ & $ 0.007 $ & 29.3 & $9.1\times10^{-5}$ & $4.4\times10^{-4}$ & $0.036$ & $0.078$ & $0.0003$ & S \\ 
HR00625 &0.0625 &$ 155 $  & $0.040 $ & $ 0.029 $ & 7.4 & $1.5\times10^{-3}$ & $-$ & $6.761$   & $3.737$& $-0.257$ & AS \\ 
HR025 &0.25 &$ 30 $ & $0.055 $ & $ 0.040 $ & 5.4 & $4.0\times10^{-3}$ & $-$ & $2.306$   & $0.934$ & $-0.384$ & AS \\ 
HR030 &0.30 &$ 24 $  & $0.056 $ & $ 0.040 $ & 5.4 & $4.5\times10^{-3}$ & $-$ & $1.955$   & $0.778$ & $-0.371$ & AS \\ 
HR035 &0.35 &$ 20 $  & $0.056 $ & $ 0.040 $ & 5.3 & $4.9\times10^{-3}$ & $-$ & $1.683$   & $0.667$ & $-0.310$ & AS \\ 
HR040 &0.40 &$ 17 $  & $0.057 $ & $ 0.040 $ & 5.3 & $5.2\times10^{-3}$ & $-$ & $1.481$   & $0.584$ & $-0.296$ & AS \\ 
HR045 &0.45 &$ 15 $  & $0.034 $ & $ 0.024 $ & 8.8 & $1.4\times10^{-3}$ & $-$ & $0.796$   & $0.519$ & $0.105$ & S \\ 
HR050 &0.50 &$ 13 $  & $0.034 $ & $ 0.024 $ & 8.9 & $1.4\times10^{-3}$ & $-$ & $0.707$   & $0.467$ & $0.114$ & S \\ 
HR055 &0.55 &$ 12 $  & $0.034 $ & $ 0.024 $ & 8.8 & $1.5\times10^{-3}$ & $-$ & $0.649$   & $0.425$ & $0.135$ & S \\ 
HR060 &0.60 &$ 10 $  & $0.034 $ & $ 0.024 $ & 8.8 & $1.6\times10^{-3}$ & $-$ & $0.593$   & $0.389$ & $0.132$ & S \\  
HR300 &3.00 &$ 0.75 $  & $0.011$ & $ 0.009 $ & 28.3 & $2.2\times10^{-4}$ & $-$ & $0.037$   & $0.078$ & $0.010$ & S \\  \hline
MS025 &0.25 &$ 30 $  & $0.030 $ & $ 0.021 $ & 10.0 & $8.4\times10^{-4}$ & $9.2\times10^{-4}$ & $1.254$   & $0.934$ & $-0.071$ & AS \\ 
MS030 &0.30 &$ 24 $  & $0.030 $ & $ 0.021 $ & 10.2 & $7.7\times10^{-4}$ & $8.5\times10^{-4}$ & $1.031$   & $0.778$ & $-0.050$ & AS \\ 
MS035 &0.35 &$ 20 $  & $0.030 $ & $ 0.022 $ & 10.0 & $8.3\times10^{-3}$ & $2.9\times10^{-4}$ & $0.896$   & $0.667$ & $0.065$ & S \\ 
HS030 &0.30 &$ 24 $  & $0.053 $ & $ 0.042  $ & 5.7 & $4.8\times10^{-3}$ & $-$ & $1.837$   & $0.778$ & $-0.370$ & AS \\ 
HS035 &0.35 &$ 20 $  & $0.049 $ & $ 0.042  $ & 6.1 & $5.0\times10^{-3}$ & $-$ & $1.462$   & $0.667$ & $-0.306$ & AS \\ 
HS040 &0.40 &$ 17 $  & $0.035 $ & $ 0.026  $ & 8.5 & $1.4\times10^{-3}$ & $-$ & $0.928$   & $0.584$ & $0.102$ & S \\ \hline
MA030 &0.30 &$ 24 $  & $0.030 $ & $ 0.021 $ & 10.1 & $7.7\times10^{-4}$ & $7.6\times10^{-4}$ & $1.034$   & $0.778$ & $-0.062$ & AS \\ 
MA035 &0.35 &$ 20 $  & $0.030 $ & $ 0.022 $ & 9.9 & $9.3\times10^{-4}$ & $4.0\times10^{-4}$ & $0.909$   & $0.667$ & $-0.017$ & AS \\ 
MA040 &0.40 &$ 17 $  & $0.030 $ & $ 0.021 $ & 10.1 & $7.8\times10^{-4}$ & $2.7\times10^{-4}$ & $0.780$   & $0.584$ & $0.059$ & S \\ 
HA040 &0.40 &$ 17 $  & $0.056 $ & $ 0.040  $ & 5.4 & $5.3\times10^{-3}$ & $-$ & $1.459$   & $0.584$ & $-0.293$ & AS \\ 
HA045 &0.45 &$ 15 $  & $0.055 $ & $ 0.040  $ & 5.4 & $5.3\times10^{-3}$ & $-$ & $1.286$   & $0.519$ & $-0.281$ & AS \\ 
HA050 &0.50 &$ 13 $  & $0.046 $ & $ 0.024 $ & 6.5 & $1.4\times10^{-3}$ & $-$ & $0.974$   & $0.467$ & $0.111$ & S
\\ \hline\hline
\end{tabular}
\label{table1}
\end{center}
\end{table*}
The simulation setup is almost the same as the model A of MYK13. 
We numerically solve a MHD convection system in a spherical shell domain defined by $(0.6R\le r \le R)$, $(0\le \theta \le \pi)$, 
and $(-\pi \le \phi < \pi)$, where $r$, $\theta,$ and $\phi$ are radius, colatitude, and longitude, respectively. 
Our model consists of two-layers, which qualitatively resemble the solar interior: stably stratified layer of thickness $0.1R$ in the range 
$(0.6R \le r \le 0.7R)$ and surrounding convective envelope of thickness $0.3R$ in $(0.7R \le r \le R)$. 

The basic equations are the fully-compressible MHD equations in the rotating frame with a constant angular velocity 
$\bm{\Omega} = \Omega_0 \bm{e}_z$ which is parallel to the coordinate axis ($\theta=0$):
\begin{eqnarray}
\frac{\partial \rho}{\partial t} & = & - \nabla\cdot \bm{f}   \;, \label{eq1} \\ 
\frac{{\partial }\bm{f}}{ \partial  t} & = & - \nabla \cdot (\bm{v} \bm{f})  - \nabla p  + \bm{j}\times\bm{B}  \nonumber \\
&& + \rho \bm{g} + 2\rho \bm{v} \times \bm{\Omega} +  \nu \left[ \nabla^2 \bm{v} + \frac{1}{3}\nabla (\nabla\cdot \bm{v})\right]  \;, \ \ \ \ \label{eq2} \\
\frac{\partial p }{\partial t} & = &  - \bm{v}\cdot\nabla p  - \gamma p \nabla\cdot \bm{v} \nonumber \\
&&+ (\gamma-1)\left[ \nabla \cdot (\kappa \nabla T) + \eta \bm{j}^2 +\Phi \right] \;, \label{eq3} \\
\frac{\partial \bm{A} }{\partial t} & = & \bm{v}\times\bm{B} - \eta \bm{j}\;, \label{eq4}
\end{eqnarray}
with 
\begin{eqnarray}
&& \Phi = 2\nu\left[ e_{ij}e_{ij} - \frac{1}{3} \left( \nabla \cdot  \bm{v} \right) \right] \;,
e_{ij} = \frac{1}{2}\left( \frac{\partial v_i}{\partial x_j} + \frac{\partial v_j}{\partial x_i}\right) \;, \nonumber \\
&& \bm{g} = -g_0/r^2\bm{e}_r\;,\  \bm{B} = \nabla \times \bm{A} \;,\ \bm{j} = \nabla\times \bm{B} \;.\nonumber
\end{eqnarray}
Here the mass density $\rho$, pressure $p$, mass flux $\bm{f} = \rho\bm{v}$, magnetic field's vector potential $\bm{A}$ are the basic 
variables. $g_0$ is the gravitational acceleration, assumed to be constant. We assume an ideal gas law $p = (\gamma -1)\rho\epsilon_i$ 
with $\gamma = 5/3$, where $\epsilon_i$ is the internal energy. The viscosity, electrical resistivity, and thermal conductivity are 
represented by $\nu$,  $\eta$, and $\kappa$ respectively. 

The initial condition is a hydrostatic equilibrium which is described by a polytropic temperature distribution with the polytropic index $m$
\begin{equation}
\frac{dT}{dr} = \frac{-g_0/r^2}{c_{\rm v}(\gamma-1)(m + 1)}\;,
\end{equation}
where $c_{\rm v}$ is the specific heat at constant volume. We choose $m=1$ and $3$ for the upper convective envelope and the lower 
stable layer, respectively. The thermal conductivity $\kappa$ is determined by requiring a constant luminosity, $L$, defined by 
$L \equiv -4\pi \kappa r^2 dT/dr$, throughout the domain.

Non-dimensional quantities are defined by setting $R = g_0=\rho_0=\mu_0=1$ where $\rho_0$ is the initial density at $r=0.6R$. The units of 
length, time, velocity, density, and magnetic field are $R$, $\sqrt{R/g_0}$, $\sqrt{g_0R}$, $\rho_0$ and $\sqrt{g_0R\mu_0\rho_0}$, 
respectively. The definitions of the Prandtl, magnetic Prandtl, and Rayleigh numbers are 
\begin{equation}
{\rm Pr}  =  \frac{\nu}{\chi},\ \ {\rm Pm} = \frac{\nu}{\eta},\ \ 
 {\rm Ra}  =  \frac{g_0d^4}{\nu\chi R^2} \left(-\frac{1}{c_p}\frac{{\rm d} s }{{\rm d}r}\right)_{r_m} \;, 
\end{equation}
where $\chi \equiv \kappa_m/(c_p\rho_{m})$ is the thermal diffusivity measured at the middle of the convection zone ($r=r_m$), 
and $d = 0.3R$ is the thickness of the convective envelope. The stratification is controlled by a normalized pressure scale height 
at the surface $\xi_{0} \equiv c_v(\gamma-1)T_s/(g_0 /R)$, where $T_s$ is the temperature at $r=R$. Here we use in all the models
$\xi_0 = 0.3$, yielding a small density contrast between top and bottom boundaries about $3$ (see Figure 1 in MYK13). This is a 
major difference between our model and actual Sun. 
 
In this paper, we focus on the two physical non-dimensional quantities, one is the ``Rossby number" (ratio of inertia to Coriolis forces) 
and the other is the ``convective Rossby number" (ratio of buoyancy to Coriolis forces) \citep[e.g.,][]{gilman77,gastine+14} which are given by 
\begin{equation}
{\rm Ro} = {\rm Co}^{-1} = \frac{v_{\rm rms} k_l}{2\Omega_0} \;, \ \ \ \ \ {\rm Ro}_{\rm conv} = \left( \frac{{\rm Ra}}{{\rm Pr} {\rm Ta}} \right)^{1/2}\;, 
\end{equation} 
respectively, where $k_l = 2\pi/d $ is the wavenumber of the largest convective eddies (this can be justified because of our weakly stratified system), 
$v_{\rm rms} \equiv [(3/2)\langle v_\theta^2 + v_r^2 \rangle]^{1/2}$ is the mean velocity, and ${\rm Ta} \equiv (2\Omega_0d^2/\nu)^2$ is the 
Taylor number. Angular brackets denote time- and volume-average ``in the convective envelope" at the saturated state. The definitions of $v_{\rm rms}$ and 
${\rm Co}$ are almost identical to those in \citet{kapyla+11,kapyla+14} and \citet{karak+14}. Note that, in this study, the convective Rossby number 
is only a function of the rotation rate, i.e., ${\rm Ro}_{\rm conv} \propto \Omega_0^{-1}$, because the other parameters are fixed in our simulations 
(see following paragraphs). 

The stress-free boundary condition for the velocity is imposed on both the radial boundaries. As for the magnetic field, we assume the 
perfect conductor at the inner boundary and the radial field condition at the outer boundary. A constant energy flux which drives the 
convective motion is imposed at the inner boundary and the temperature is fixed to be $T_s$ at the outer boundary. 

The equations~(\ref{eq1})--(\ref{eq4}) are solved by the second-order central difference scheme with spatial discretization with Yin-Yang grid. 
See \citet{kageyama+04} and MYK13 for details. Non-dimensional parameters $\rm{Pr}=0.27$, $\rm{Pm}=4.0$, and $\rm{Ra} = 2.2\times10^5$ 
(i.e., $\nu = 7.8\times 10^{-5}$, $\eta = 2.0 \times 10^{-5}$, and $\kappa = 3.9\times 10^{-4}$) are commonly adopted in all the 
models studied here. The total grid size for all the simulation runs is $81$ (in $r$) $\times 131$ (in $\theta$) $\times 393$ (in $\phi$) $\times 2$ 
(Yin \& Yang). To explore the response of the large-scale flow to the rotation rate, we vary the magnitude of $\Omega_0$ in the MHD and 
HD models while keeping the background hydrostatic state unchanged. A random temperature perturbation and small ``seed'' magnetic 
field (only for MHD models) are introduced at the same time in the convection zone when the calculation starts.
\begin{figure*}[htpb]
\begin{center}
\includegraphics[width=17cm,clip]{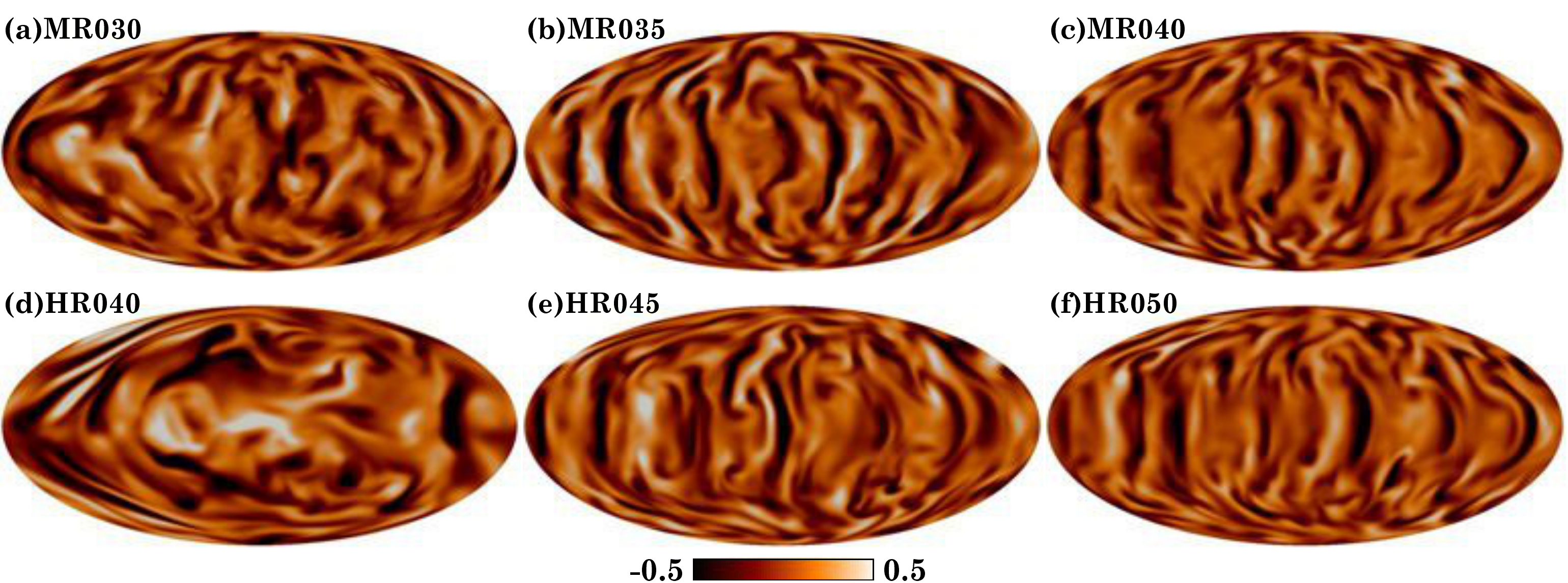}
\caption{The radial velocity profile $v_r (r,\theta)$ at $r=0.85R$ when $t \simeq 360\tau_c$ on a spherical surface. 
The normalization unit is the mean velocity $v_{\rm rms}$ of each model. The panels (a)--(c) 
correspond to the MHD models with $\Omega_0=0.3$, $0.35$ and $0.4$, and the panels (d)--(f) are for the HD models with 
$\Omega_0=0.4$, $0.45$ and $0.5$. The lighter and darker tones depict upflow and downflow velocities.  }
\label{fig1}
\end{center}
\end{figure*}
\section{Numerical Results}
Our simulation models are summarized in Table 1. Model names are given in the left most column. The first letter in the model name denotes 
the presence and absence of the magnetic field. The labels of MHD models contain the letter M. HD models have a label beginning with H. 
The second letter denotes the initial rotation profile. The labels of basic runs with the initial rigid rotation contains the letter R. Models started 
with a solar-type or an anti-solar type DR have a letter S or A. The number indicates the given rotation rate $\Omega_0$ with the decimal point 
removed. 

The mean quantities evaluated at the saturated state ($v_{\rm rms}$, $B_{\rm eq}$, $\tau_c$, $\epsilon_{\rm K}$, and $\epsilon_{\rm K}$), 
the Rossby number (${\rm Ro}$), and the convective Rossby number (${\rm Ro}_{\rm conv}$) are shown in Table 1, where 
$\tau_c \equiv d/v_{\rm rms }$ is the convective turn-over time, $B_{\rm eq} \equiv \langle (3/2)\rho\; (v_\theta^2 + v_r^2) \rangle^{1/2}$
is the the equipartition strength of magnetic field, $\epsilon_{\rm K} \equiv \langle \rho \bm{v}^2 \rangle_{\rm fs}/2$ and 
$\epsilon_{\rm M} \equiv \langle \bm{B}^2 \rangle_{\rm fs}/2$ are the mean kinetic and magnetic energy densities. Angular brackets with 
subscript ``{\rm fs}" denote time- and volume-average in the full-spherical shell domain at the saturated state. As a measure of the convection 
strength, the estimated super-criticality level, defined by $\delta = ({\rm Ra} - {\rm Ra}_c)/{\rm Ra}_c$, is shown in the second column, where 
${\rm Ra}_c$ is the critical Rayleigh number (see caption for the definition of ${\rm Ra}_c$).  In the second last and last columns, 
DR parameter $\alpha_e$ (see \S3.2 for the definition) and, the resultant DR profile, S (solar-type) or AS (anti-solar type), are summarized. 

We perform three sets of runs. In the first set, we run models from the rigid rotation (Set R: Runs MR and HR) 
with the initial conditions described in \S2. In the second and third sets, we examine the bistability of the rotation profile by taking either a solar-type 
(Set S: Runs MS and HS) and an anti-solar type (Set A: Runs MA and HA) solution as initial conditions. The MHD effects on the 
convective flows are studied in each set. 
\subsection{Convective Flows in Basic Runs with Initial Rigid Rotation}
\begin{figure*}[htbp]
\begin{center}
\includegraphics[width=17cm,clip]{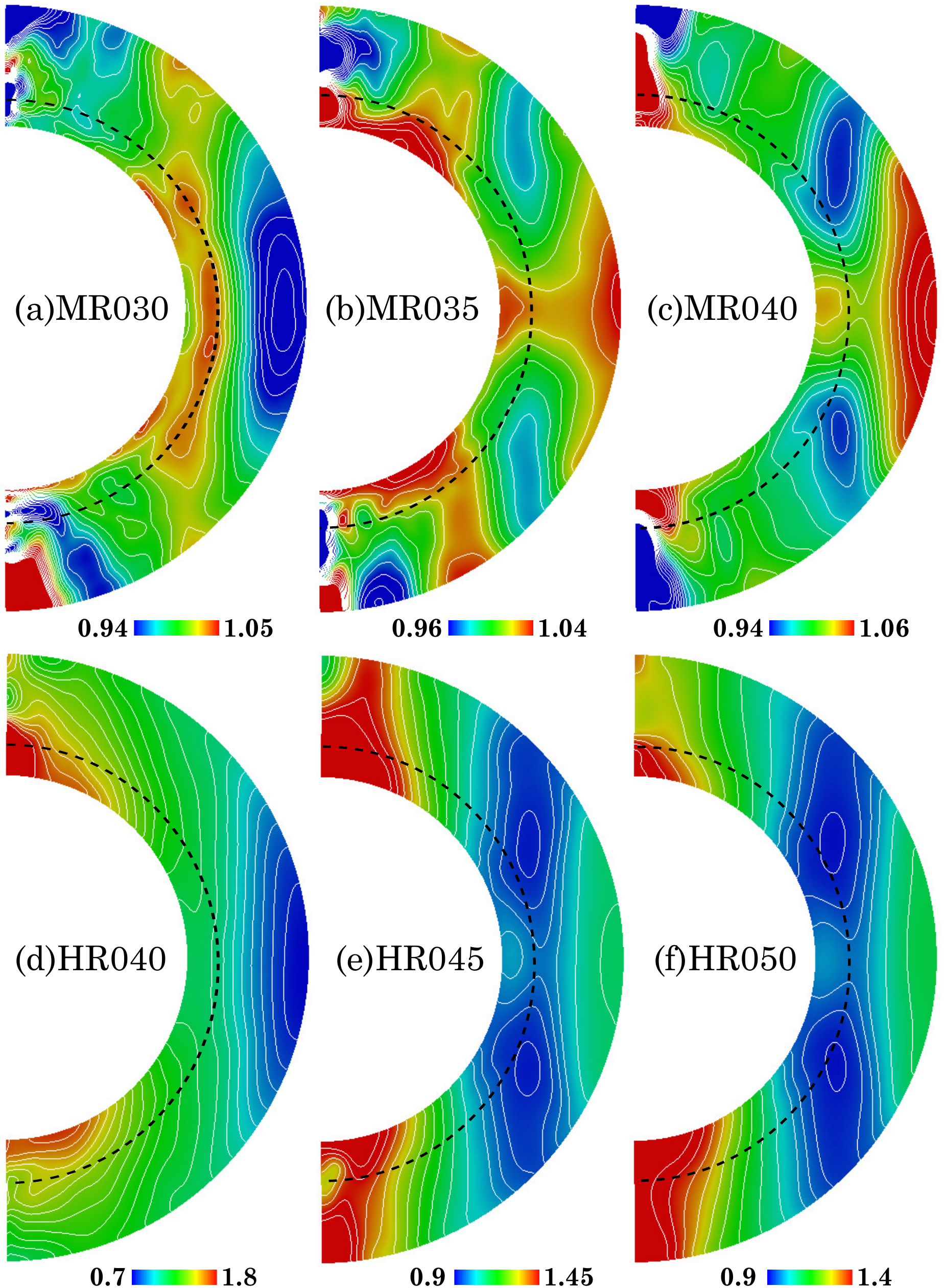}
\caption{Time-averaged mean angular velocity $\bar{\Omega}(r,\theta)$. The panels (a)--(c) correspond to the MHD models with 
$\Omega_0=0.3$, $0.35$ and $0.4$, and the panels (d)--(f) are for the HD models with $\Omega_0=0.4$, $0.45$ and $0.5$. The 
white solid line denotes the iso-rotation contour. The black dashed line denotes the interface between convective and radiative zones. }
\label{fig2}
\end{center}
\end{figure*}
After the convective motion sets in, it reaches a saturated state before $\simeq 100\tau_c$ in the basic run with the initial rigid rotation. We 
have run the simulations up to around $400 \tau_c$, which is almost comparable to the magnetic diffusion time, and studied the properties 
of the convective flow at the equilibrated state. 

Figure 1 shows, in the Mollweide projection, the radial velocity distribution when $t \simeq 360\tau_c$ on spherical surface at $r = 0.85R$. 
The normalization unit is the mean velocity, $v_{\rm rms}$, of each model. 
Panels (a)--(c) correspond to the MHD models (MR030, MR040, and MR045) and panels (d)--(f) are for the HD models (HR040, HR045, 
and HR050). The lighter and darker tones depict upflow and downflow velocities in the range $|v_r/v_{\rm rms}| \le 0.5$. 

The rotating stratified convection is characterized by up-down asymmetry in all the models: the slower upflow dominant cells surrounded 
by the faster and narrower downflow lanes \citep[e.g.,][]{spruit+90,miesch05}. As the rotation rate increases, the convective cell shrinks 
and its structure is changed from the cellular pattern to the elongated columnar pattern formed in the equatorial region 
\citep[e.g.,][]{busse70,glatzmaier+81,brummell+96, brummell+98, miesch+00}. The transition from the cellular to columnar convections  
has been obsered in earlier simulations \citep{kapyla+11, kapyla+14, gastine+13, gastine+14, guerrero+13} and occurs at around 
$\Omega_0 = 0.35$ (${\rm Ro}_{\rm conv} = 0.67$) in the MHD model but $\Omega_0 = 0.45$ (${\rm Ro}_{\rm conv} = 0.52$) in the HD 
model. The magnetic field affects the convective motion, suggesting its impacts on the turbulent angular momentum transport and thus the 
resultant large-scale DR. 

Shown in Figure 2 is the profile of the mean DR, defined by $\bar{\Omega}(r,\theta)=\bar{v}_{\phi}/(r \sin{\theta}) + \Omega_0$, where the 
overbar denotes the time- and azimuthal-average. The normalization unit is $\Omega_0$ of each model. The time average spans in the 
range of $200 \lesssim t/\tau_c \lesssim 300$. Panels (a)--(c) correspond to the MHD models and panels (d)--(f) are the HD models. The 
red (blue) tone denotes the higher (lower) angular velocity. The white lines are iso-rotation contours. The interface between the convective 
envelope and the radiative layer is denoted by black dashed line. 

As shown in earlier studies, there exists two regimes in the rotation profile: an anti-solar type DR with slower equator [(a) \& (d)] and 
a solar-type DR with faster equator [(b),(c),(e) \& (f)] \citep[e.g.,][]{gilman77, kapyla+11, kapyla+14, gastine+13, gastine+14, guerrero+13}. 
While the anti-solar type DR is established in the model with the smaller rotation rate, the solar-type DR develops in the regime of the 
higher rotation rate. The most remarkable difference between the MHD and HD models is the critical rotation rate (and thus the critical 
convective Rossby number) for the transition. While the transition between the anti-solar and solar-type DR profiles occurs at around 
$\Omega_0 = 0.35$ (${\rm Ro}_{\rm conv} = 0.67$) in the MHD model, it does at around $\Omega_0 = 0.45$ (${\rm Ro}_{\rm conv} = 0.52$) 
in the HD model. 

The DR profile is dominated by so-called ``Taylor-Proudman balance" both in the MHD and HD models \citep[e.g.,][]{pedlosky87} and is still 
far from the solar conical isotachs deduced from the helioseismic measurement \citep[e.g.,][]{kitchatinov+95,miesch05,rempel05,masada11}. 
Nevertheless, it is intriguing that the iso-rotation contour is more conical in the MHD model than in the HD model. Especially, the radial shape 
of the isotachs is pronounced around the equator in the MHD model near the transition (see MR035). 

When comparing the two models at the same rotation rate, we can see that the rotational shear is generally weaker in the MHD model. A 
strong polar jet (polar vortex) appeared in the HD model is also suppressed in the MHD model (see HR045 and HR050). In addition, the 
MHD effect seems to suppress the (viscous) spreading of the rotation profile of the convective envelope into the radiative layer in all the 
MHD models, as was observed in \S4 of MYK13 \citep[see, e.g.,][]{spiegel+92,gough+98}. The reduction of the rotational shear by the 
influence of the magnetic fields agrees with that has been reported in the earlier studies \citep[e.g.,][]{brun+04,beaudoin+13,fan+14,karak+14} 
and is discussed further in \S3.2. 

Figure 3 shows the time-averaged mean meridional flows. Panels (a)--(c) correspond to the MHD models and panels (d)--(f) are the HD 
models around the transition between the solar and anti-solar type rotation profiles. The color contour depicts the meridional flow velocity, 
defined by $v_{m} = (\bar{v}_r^2 + \bar{v}_\theta^2)^{1/2}$. The normalization unit is $v_{\rm rms}$ of each model. The streamlines are 
over-plotted with a length proportional to the flow speed. The white dashed line denotes the interface between the convective and radiative layers. 

The circulation flow is primarily counter-clockwise (clockwise) in the bulk of the convection zone in the northern (southern) hemisphere. 
This is consistent with the observations of the solar sub-surface meridional flow \citep[e.g.,][]{komm+93,giles+97,ulrich10,zhao+13}, and 
a common feature of all the models. However, there is a difference in the circulation pattern. While a large single-cell is formed in the 
model with the anti-solar type DR (see MR030 and HR040), the model with the solar-type DR shows a multiple-cell pattern 
(see MR040 and HR050). A similar transition from single to multiple cell patterns of the meridional flow has been observed before in 
different settings \citep[e.g.,][]{kapyla+11, kapyla+14, gastine+13, karak+14}. We note that, as well as the rotation pattern, the transition 
of the meridional flow pattern occurs at lower $\Omega_0$ (thus higher ${\rm Ro}_{\rm conv}$) in the MHD model than in the HD model. 
\begin{figure*}[htbp]
\begin{center}
\includegraphics[width=17cm,clip]{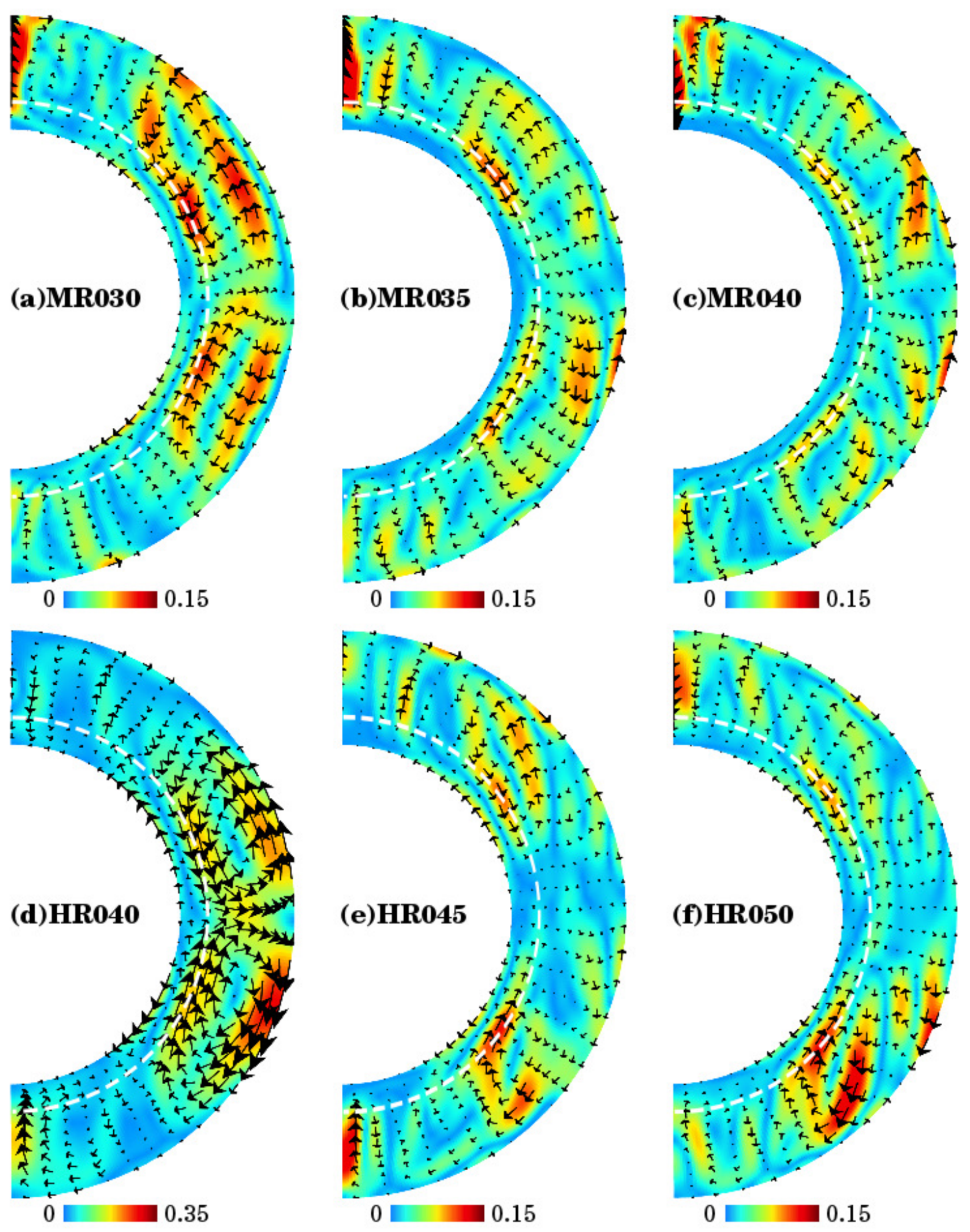}
\caption{Time-averaged mean meridional flows. Panels (a)--(c) correspond to the MHD models (MR030, MR035, MR040) and 
panels (d)--(f) are the HD models (HR040, HR045, HR050). The color contour depicts the meridional flow velocity, defined by
$v_{m} = [\bar{v}_r^2 + \bar{v}_\theta^2]^{1/2}$ normalized by $v_{\rm rms}$ of each model. The streamlines are overplotted 
with a length proportional to the flow speed. The white dashed line denotes the interface between convection and radiative zones.}
\label{fig3}
\end{center}
\end{figure*}
\subsection{Control Parameter of Differential Rotation} 
The transition between the anti-solar and solar-type DRs is a result of the structural difference of the whole three-dimensional convective 
motion separated at the critical rotation rate. Here we quantitatively examine the dependences of the DR profiles on two key 
diagnostic parameters, the convective Rossby number ${\rm Ro}_{\rm conv}$ and the Rossby number ${\rm Ro}$. Then we address which 
can better capture the transition between the anti-solar type and solar-type DRs. Following \citet{gastine+14}, we quantify the DR by the 
amplitude of the mean azimuthal flow at the equatorial surface:
\begin{equation}
\alpha_e = \frac{\bar{v}_\phi (r = R,\theta = \pi/2)}{\Omega_0 R} \;.
\end{equation}
The positive and negative $\alpha_e $ denote the solar-type and anti-solar type DRs. 

Figure 4(a) shows the DR parameter ($\alpha_e$) as functions of the convective Rossby number (${\rm Ro}_{\rm conv}$) [bottom axis] 
and the corresponding rotation rate ($\Omega_0$) [top axis]. The blue crosses and red diamonds denote the MHD and HD models. The 
horizontal and vertical dashed lines indicate $\alpha_e = 0$ and ${\rm Ro}_{\rm conv} = 1$.  As stated in \S~3.1, the DR profile transits 
from the solar to anti-solar type regimes as the convective Rossby number increases. The absolute amplitude of $\alpha_e$ is, on average, 
larger in the anti-solar type DR than in the solar-type DR, and takes the   maximum and minimum around the transition. These are common 
features of the two models and are compatible with earlier studies \citep{gastine+14,kapyla+14, karak+14}. The critical value of 
${\rm Ro}_{\rm conv}$ for the transition is different in the two models. It occurs when $0.67 < {\rm Ro}_{\rm conv} < 0.78$ for the MHD 
model and $0.52 < {\rm Ro}_{\rm conv} < 0.58$ for the HD model, indicating a more gradual transition under the influence of the magnetic 
field. When comparing two models at the same ${\rm Ro}_{\rm conv}$, the magnitude of $\alpha_e$ is smaller in the MHD model than in 
the HD model. We confirm that the magnetic field has a role in suppressing the DR and helps to produce solar-type DR, which have been 
suggested in MYK13, \citet{fan+14} and \citet{karak+14}. 

Our intriguing finding is that the critical value of the Rossby number for the transition is almost the same between the MHD and HD models. 
Shown in Figure 4(b) is the dependence of the DR parameter on the   Rossby number (${\rm Ro}$). The blue crosses and red diamonds 
are corresponding to the MHD and HD models. The horizontal and vertical dashed lines denote $\alpha_e = 0$ and ${\rm Ro} = 1$. The DR 
parameter decreases with the increase of the Rossby number in both models around the transition. The transition between the solar and 
anti-solar type DR profiles occurs at ${\rm Ro} \simeq 1.0$ both in the MHD and HD models. Not only the critical value, the sharpness of the 
transition is also similar between the two models. The critical Rossby number for the transition obtained here gives a good agreement with 
that obtained in the hydrodynamic simulations by \citet{kapyla+11,kapyla+14}. These indicate that the transition between the two rotation 
regimes is controlled by the Rossby number rather than by the convective Rossby number (i.e., the rotation rate itself in our study) regardless 
of the presence of the magnetic field. 

We stress here that \citet{karak+14} also studied the magnetic influence on the DR profile by comparing HD and MHD models, but 
did not discuss their ${\rm Ro}$-dependences. Reading the critical Rossby number for the transition from Table~1 of \citet{karak+14} and \citet{kapyla+14}, 
it is found that the transition occurs at a similar value of ${\rm Ro}$ in their MHD and HD models as well . Our finding is thus consistent with 
the earlier studies. The influence of the magnetic field on the DR profiles was also briefly mentioned in \citet{gastine+14}. However, they focused on 
the dependence of the DR profile on the ``local Rossby number (${\rm Ro}_l$)" and did not examine its ${\rm Ro}$-dependence. Here the local 
Rossby number is defined by ${\rm Ro}_l = \bar{l}_u \delta u_{\rm rms} /(\pi \Omega_0 d) $, where $\delta u_{\rm rms}$ is the RMS value of the 
non-axisymmetric (fluctuating) component of the velocity (including both meridional and azimuthal components) and $\bar{l}_u$ is the  mean 
spherical harmonic degree obtained from the kinetic energy spectrum \citep[e.g.,][]{christensen+06,schrinner+12}. 
Since they showed that the critical ``local" Rossby number for the transition is different between MHD and HD models, we can say 
that the Rossby number is even better than the local Rossby number at least to describe the transition between the anti-solar and solar-type 
DR profiles in a unified manner. 

In order to reveal a cause of the difference in the critical value of the convective Rossby number between the two models, we present, 
in Figure 5, the dependence of the mean convective velocity ($v_{\rm rms}$) on the convective Rossby number for the MHD (blue crosses) 
and HD (red diamonds) models. Here the regime around the transition is focused. When comparing two models at the same 
${\rm Ro}_{\rm conv}$, the MHD model has the smaller convective velocity than the HD model. The MHD model thus can maintain the 
solar-type DR even at the lower rotation rate (thus higher convective Rossby number) in comparison with the HD model. Since the Lorentz 
force of the dynamo-generated magnetic field puts the brakes on the convective motion \citep[e.g.,][]{cattaneo+03,brun+04}, the MHD model 
becomes more rotationally-constrained than the HD model. This could be the reason why the magnetic field helps to produce the solar-type 
DR. \citet{fan+14} and \citet{karak+14} also argued that the reduction of the convective velocity by the dynamo-generated magnetic field is 
responsible for the difference of the critical rotation rate (i.e., critical convective Rossby number) for the solar to anti-solar transition between 
the HD and MHD models. It is interesting that such a drastic difference between the MHD and HD models observed in $v_{\rm rms}$ 
(Fig.5) is unified in the view of the ${\rm Ro}$-dependence (Fig.4b). 
\begin{figure*}[htpb]
\begin{center}
\includegraphics[width=18cm,clip]{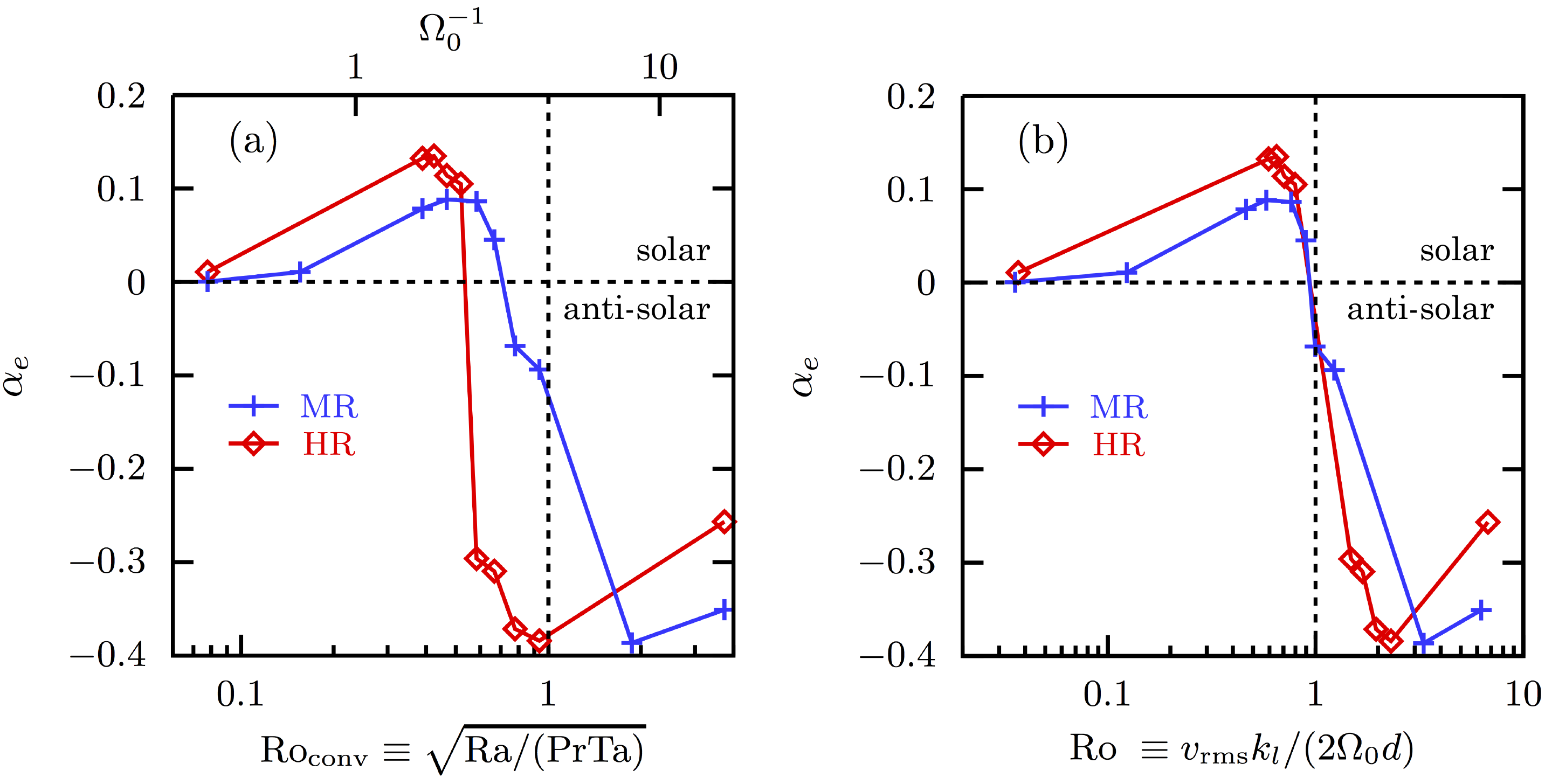}
\caption{(a) DR parameter $\alpha_e$ as functions of the convective Rossby number (${\rm Ro}_{\rm conv}$) and 
the corresponding rotation rate ($\Omega_0$). The horizontal and vertical dashed lines indicate $\alpha_e = 0$ and 
${\rm Ro}_{\rm conv} = 1$. (b) DR parameter $\alpha_e$ as a function of the   Rossby number (${\rm Ro}$). 
The horizontal and vertical dashed lines indicate $\alpha_e = 0$ and ${\rm Ro} = 1$. The blue crosses and red 
diamonds denote the MHD and HD models with the initial rigid rotation in both panels. }
\label{fig4}
\end{center}
\end{figure*}
\begin{figure}[htpb]
\begin{center}
\includegraphics[width=8.5cm,clip]{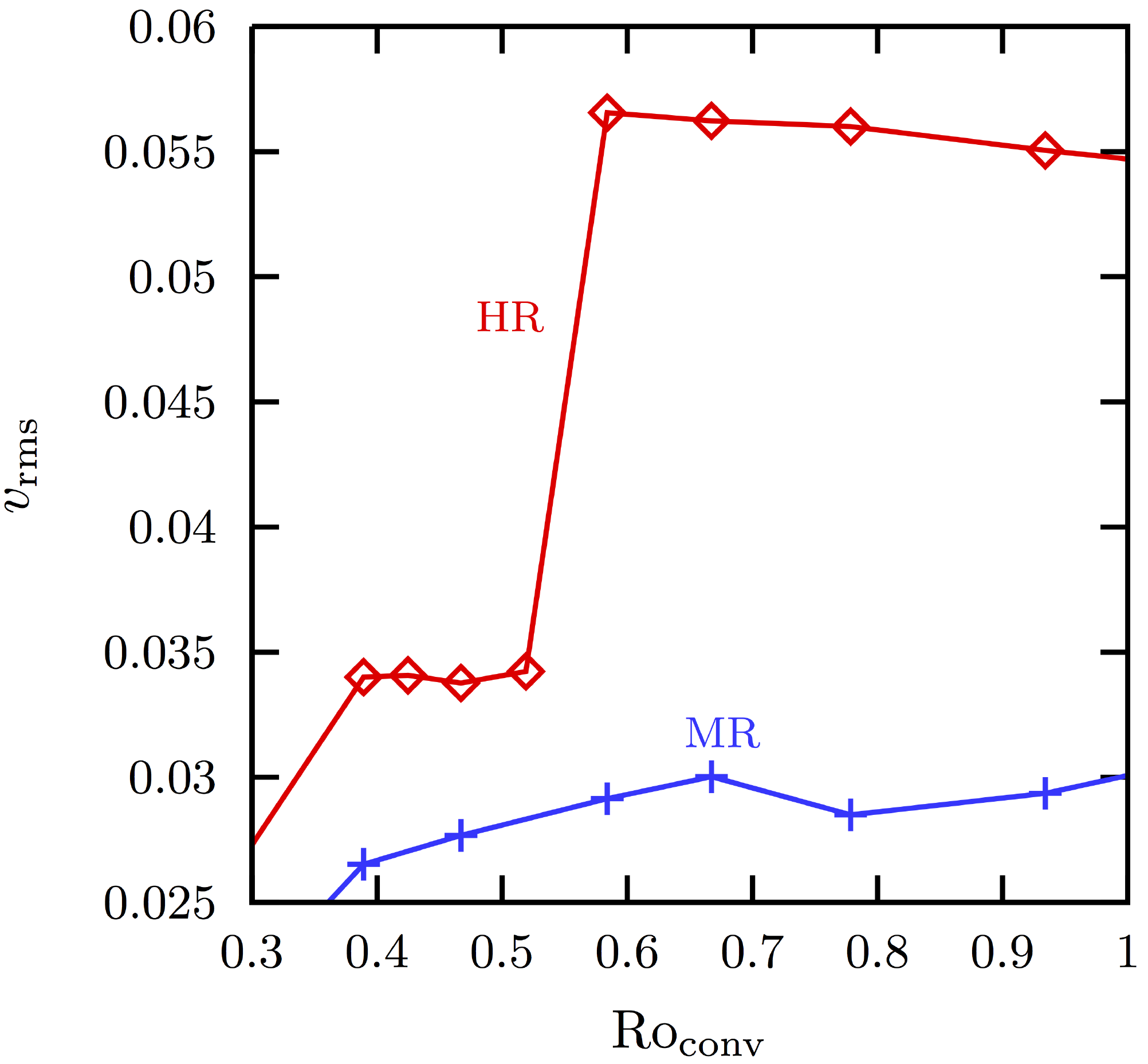}
\caption{Mean convective velocity $v_{\rm rms}$ as a function of the convective Rossby number (${\rm Ro}_{\rm conv}$). 
The blue crosses and red diamonds denote the MHD and HD models with the initial rigid rotation.}
\label{fig5}
\end{center}
\end{figure}
\subsection{Bistability of the Rotation Profile}
The transition between the solar and anti-solar type rotation patterns has been known since \citet{gilman77} and studied in various settings 
\citep{kapyla+11,gastine+13,guerrero+13}. However, the bistability of the rotation profile has not been sufficiently-studied because it has 
only recently been discovered \citep{gastine+14,kapyla+14}. We revisit here the dependence of the rotation profile on the history both in 
the MHD and HD models by taking either a solar-type (Set S: Runs MS and HS) and an anti-solar type (Set A: Runs MA and HA) solution 
as initial conditions. 

In the Set S, we perform simulations by starting from the saturated state of MR040 (HR050) with the solar-type DR for the MHD (HD) run 
and decrease the rotation rate. In contrast, in the Set A, we choose the final state of MR025 (HR030) with the anti-solar type DR as the 
progenitor for the MHD (HD) model and increase the rotation rate. The DR parameter ($\alpha_e$) for these models are shown in 
Figure 6(a) as a function of the convective Rossby number (${\rm Ro}_{\rm conv}$). The models MS, MA, HS and HA are denoted by cyan triangles, green circles, 
pink crosses and orange squares, respectively. The DR parameters of the basic runs (MR and HR) are also shown by blue crosses and red 
diamonds as references. 

As shown in \citet{gastine+14} and \citet{kapyla+14}, the bistability of the rotation profile can be observed near the transition in the HD 
model. In the Set HS, the transition of the HD solution occurs when $0.58 < {\rm Ro}_{\rm conv} < 0.67$ while it does when 
$0.47 < {\rm Ro}_{\rm conv} < 0.52$ in the Set HA. On the other hand, the set MS started 
from the solar-type DR reproduces the similar results with the set MR with the initial rigid rotation (see cyan triangles and blue crosses). 
This is consistent with the recent studies of \citet{fan+14} and \citet{karak+14} who found in their convective dynamo simulations that 
the bistability is disappeared and the DR profile becomes independent from the hysteresis when allowing the growth of the MHD dynamo. 
However, in the set MA which is started from the anti-solar type DR, the bistability seems to be not disappeared (see green circles). 

The remaining of the bistability in the set MA of our MHD models may be related to the presence of the stable layer underlying the convective 
envelope, which is a difference between our simulation model and the models adopted in \citet{fan+14} and \citet{karak+14}. Carefully 
studying the rotation profile of the MR030 (Fig.2a) which has a very similar rotation profile with the progenitor for the set MA, we can find 
that the stable layer has a strong equatorial acceleration despite the anti-solar type DR established in the convection zone. Since the differential 
rotation in the stable layer should change the latitudinal temperature and entropy distributions to achieve the thermal wind balance there 
\citep[e.g.,][]{kitchatinov+95,rempel05}, it may enforce the anti-solar type DR even in the high $\Omega_0$ regime which 
should provide the solar-type DRs. Our speculation on the role of the stable layer in the maintenance of the internal rotation will be 
quantitatively examined in a subsequent paper. 

It is interesting that, by using the Rossby number (${\rm Ro}$), we can again unify the transition between the solar and anti-solar type rotation profiles 
independently of the hysteresis. Shown in Figure 6(b) is the DR parameters of all the models as a 
function of the   Rossby number. The symbols have the same meanings as those in Figure 6(a). It is found that the transition occurs 
at ${\rm Ro} \simeq 1$ in all the simulation sets. Not only the critical value, the sharpness of the transition is also similar between all the 
cases. This confirms that the transition between the two rotation regimes is better captured by the Rossby number rather than by 
the convective Rossby number. 

One important finding in our study is that the initial rotation profile, i.e., the evolution history of the stellar rotation, has an impact on the 
convective velocity if we does not allow the growth of the magnetic field. The dependence of the mean convective velocity ($v_{\rm rms}$) 
on the convective Rossby number (${\rm Ro}_{\rm conv}$) is shown in Figure 7. The symbols have the same meanings as those in Figure 6(a). 
Here the regime around the transition is focused. The bistable nature of the convective velocity can be seen in the HD models, while the MHD models have 
similar convective velocity around the transition. Thus, in the HD cases, the Set S (Set A) with the initial solar-type (anti-solar type) DR 
becomes more rotationally-dominated (inertia dominated) than the other sets. In contrast, in the MHD cases, the evolution history of the 
DR does not change the convective velocity and the rotational dominance in the system. This would be due to the regulation of the 
convective velocity by the dynamo-generated magnetic field. These indicate that the bistability appeared in the rotation profile is essentially 
a consequence of the hysteresis of the convective velocity, which can be weakened by the dynamo-generated magnetic field.

Since the convective Rossby number denotes the relative importance of the buoyancy force in comparison with the Coriolis force, the 
magnetic effect and the evolution history are not directly reflected in its definition. In contrast, the   Rossby number describes the significance 
of the inertia force arising as the results of the hysteresis and the non-linear interaction between the convective motion and the magnetic 
field, relative to the Coriolis force. Therefore, the effects of the magnetic field and the hysteresis are naturally reflected in it. This would be the 
reason why the transition between the solar and anti-solar type rotation profiles can be unified in the view of the ${\rm Ro}$-dependence. 
\begin{figure*}[htpb]
\begin{center}
\includegraphics[width=18cm,clip]{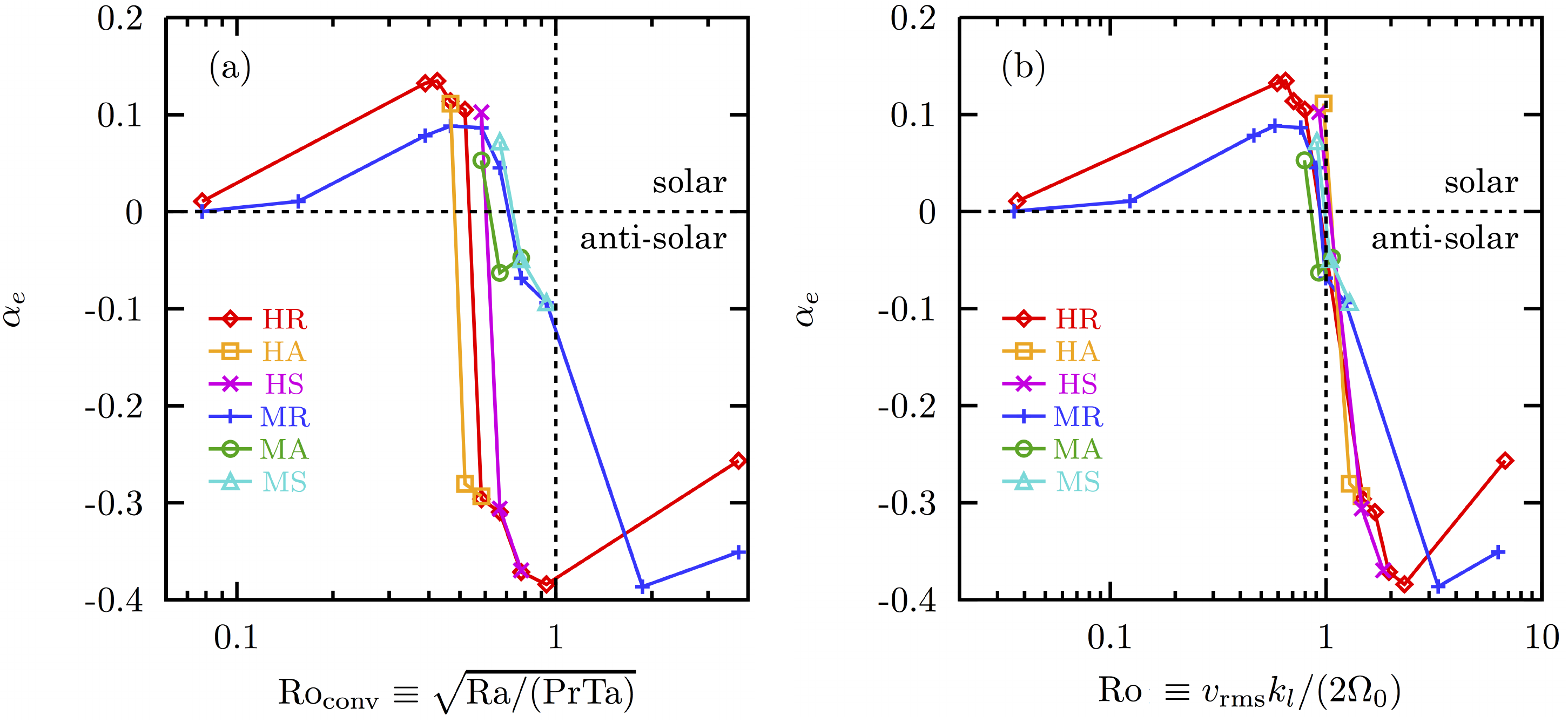}
\caption{DR parameters ($\alpha_e$) of the models with non-rigid initial rotations as functions of (a) the convective Rossby number 
(${\rm Ro}_{\rm conv}$) and (b) the   Rossby number (${\rm Ro}$). The models MS, MA, HS and HA are denoted by cyan triangles, 
green circles, pink crosses and orange squares, respectively in both panels. The horizontal and vertical dashed lines indicate 
(a) $\alpha_e = 0$ \& ${\rm Ro}_{\rm conv} = 1$ and (b) $\alpha_e = 0$ and ${\rm Ro} = 1$. The DR parameters of the basic runs 
(MR and HR) are shown by blue crosses and red diamonds as references. }
\label{fig6}
\end{center}
\end{figure*}
\begin{figure}[htpb]
\begin{center}
\includegraphics[width=8.5cm,clip]{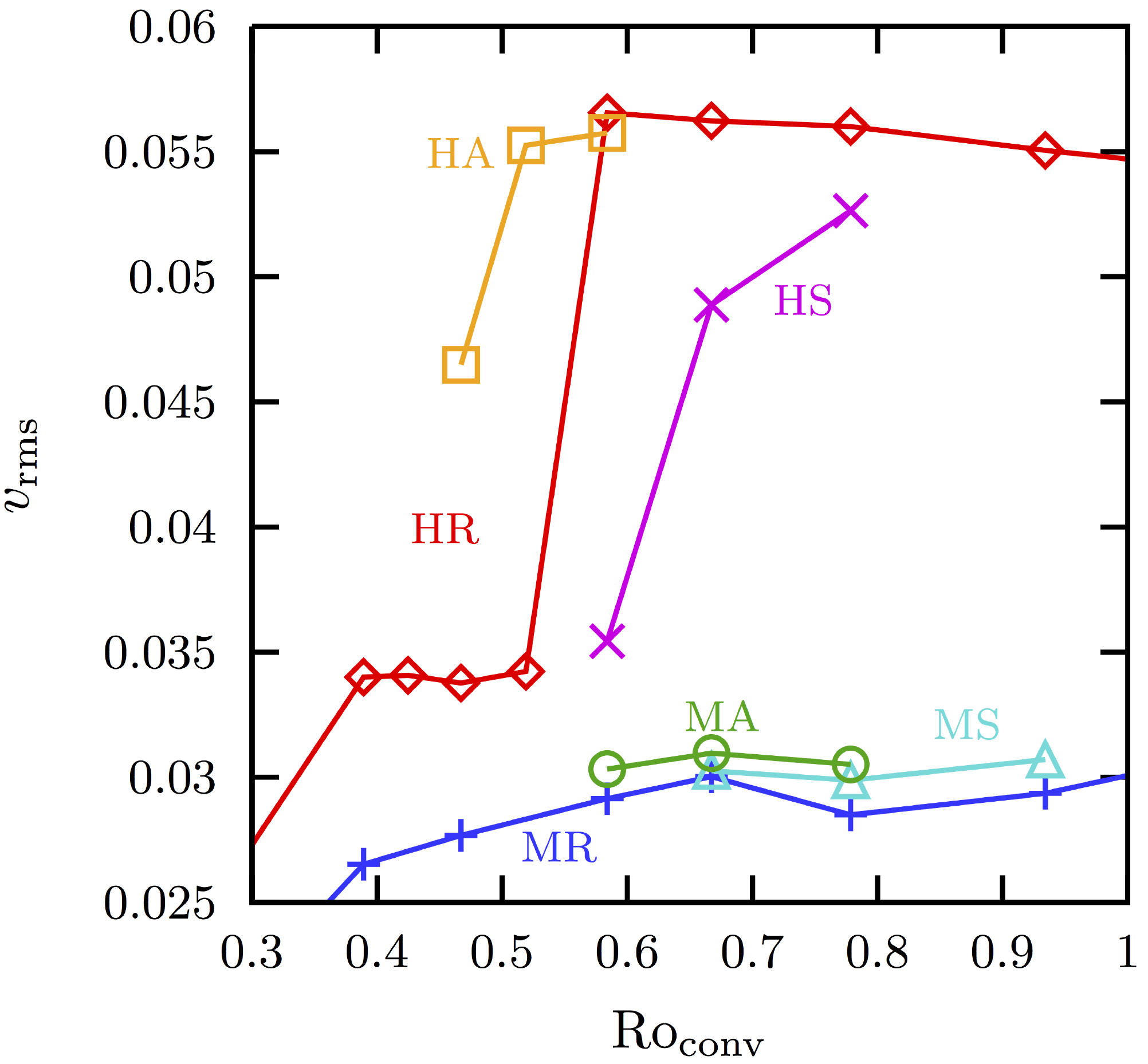}
\caption{Mean convective velocity $v_{\rm rms}$ as a function of the convective Rossby number (${\rm Ro}_{\rm conv}$). 
The models MR, MS, MA, HR, HS and HA are denoted by blue crosses, cyan triangles, green circles, red diamonds, 
pink crosses and orange squares, respectively.}
\label{fig7}
\end{center}
\end{figure}
\subsection{Dependence of Magnetic Dynamo Activity on Rotation Rate}
The MHD model exhibits a rich variety of the magnetic dynamo activity depending on the rotation rate. 
We show the time-latitude diagram of (a)$\langle B_r \rangle_{\phi}$ (left column), (b)$\langle B_\theta \rangle_{\phi}$ (middle column) 
and (c)$\langle B_\phi \rangle_{\phi}$ (right column) of MR00625, 025, 030, 050, 060, and 150 at $r = 0.65R$ (mid-stable zone) in Figure 8 
and at $r = 0.85R$ (mid-convection zone) in Figure 9. Here the angular brackets with subscript $\phi $ denote the azimuthal average at a 
given depth. The time and amplitude are normalized by $\tau_c$ and $B_{\rm eq}$, respectively. The red and blue tones denote the 
positive and negative mean-field strength. The horizontal dashed-line in each panel denotes the equator. The upper three models have the 
anti-solar type DR and the lower three models are characterized by the solar-type DR. MR025--060 are the models near the transition. 

The model with the smallest rotation rate is dominated by the turbulent magnetic field and has little large-scale component both in the 
radiative and convective layers (see MR00625). This would be due to the weak Coriolis force, yielding weak differential rotation. At around the 
transition of the rotation profile, the spatiotemporal coherence of the magnetic field becomes remarkable. In the slow rotation regime with 
the anti-solar type DR, a quasi-steady large-scale magnetic field with a dipole symmetry is organized (see MR025 and 030). $\langle B_r \rangle_{\phi}$ 
and $\langle B_\theta \rangle_{\phi}$ are stronger in the convection zone rather than in the radiative zone, while $\langle B_\phi \rangle_{\phi}$ 
is stronger in the stable zone than in the convection zone. Among the three magnetic components, $\langle B_\phi \rangle_{\phi}$ is dominant 
in these models. The large-scale $B_\phi$-component is the strongest at the mid-latitude and is antisymmetric about the equator. This is 
consistent with the shearing of the $B_\theta$-component, i.e., $\Omega$-process, by the anti-solar type DR. 

The properties of the magnetic dynamo change drastically across the transition of the DR profile. The large-scale magnetic field shows the polarity 
reversal in the fast rotation regime with the solar-type DR. In the models near the transition, i.e., MR050 and 060, the large-scale magnetic component 
with polarity reversals is built up mainly in the stable zone. It has commonly $B_\phi$-dominance. The large-scale $B_\phi$-component is the strongest 
at around the equator in these models unlike the models with the anti-solar type DR in which it is the strongest at the mid-latitude. 
In MR050, the convection zone is dominated by the turbulent magnetic field. However, we can observe a weak spatiotemporal coherence of the magnetic 
field in the convection zone in MR060. It is interesting that the cycle of the polarity reversal of the large-scale magnetic field around the equator is shorter 
in the convection zone than in the radiative zone. 

The most intriguing dynamo activity can be observed in the model with a fast rotation beyond the transition (see MR150). Note that this 
model has 3--4 times higher rotation rate than that adopted in the model around the transition. In the stable zone, the strong large-scale 
magnetic component with a dipole symmetry is built up. It concentrates at the high latitudinal region ($\simeq 80^\circ$) and seems to 
have a long cycle period of the polarity reversal ($\tau_{\rm cyc} \gtrsim 200\tau_{c}$). The spatial distribution of the $B_\phi$-field 
suggests that the $\Omega$-effect plays a crucial role in amplifying the large-scale magnetic field in the stable region. In the convective  
envelope, two types of the dynamo mode can be clearly observed. The long-lasting magnetic component with a dipole symmetry, which 
is similar to that in the stable zone, is found in the high latitude. In contrast, at around the equatorial region, the oscillatory dynamo mode 
with a shorter cycle period ($\tau_{\rm cyc} \simeq 10\tau_c$) and a poleward migration is excited and sustained. The difference of the 
dynamo mechanism between the models with the different rotation profiles is discussed in \S4. 
\begin{figure*}[tpb]
\begin{center}
\includegraphics[width=18cm,clip]{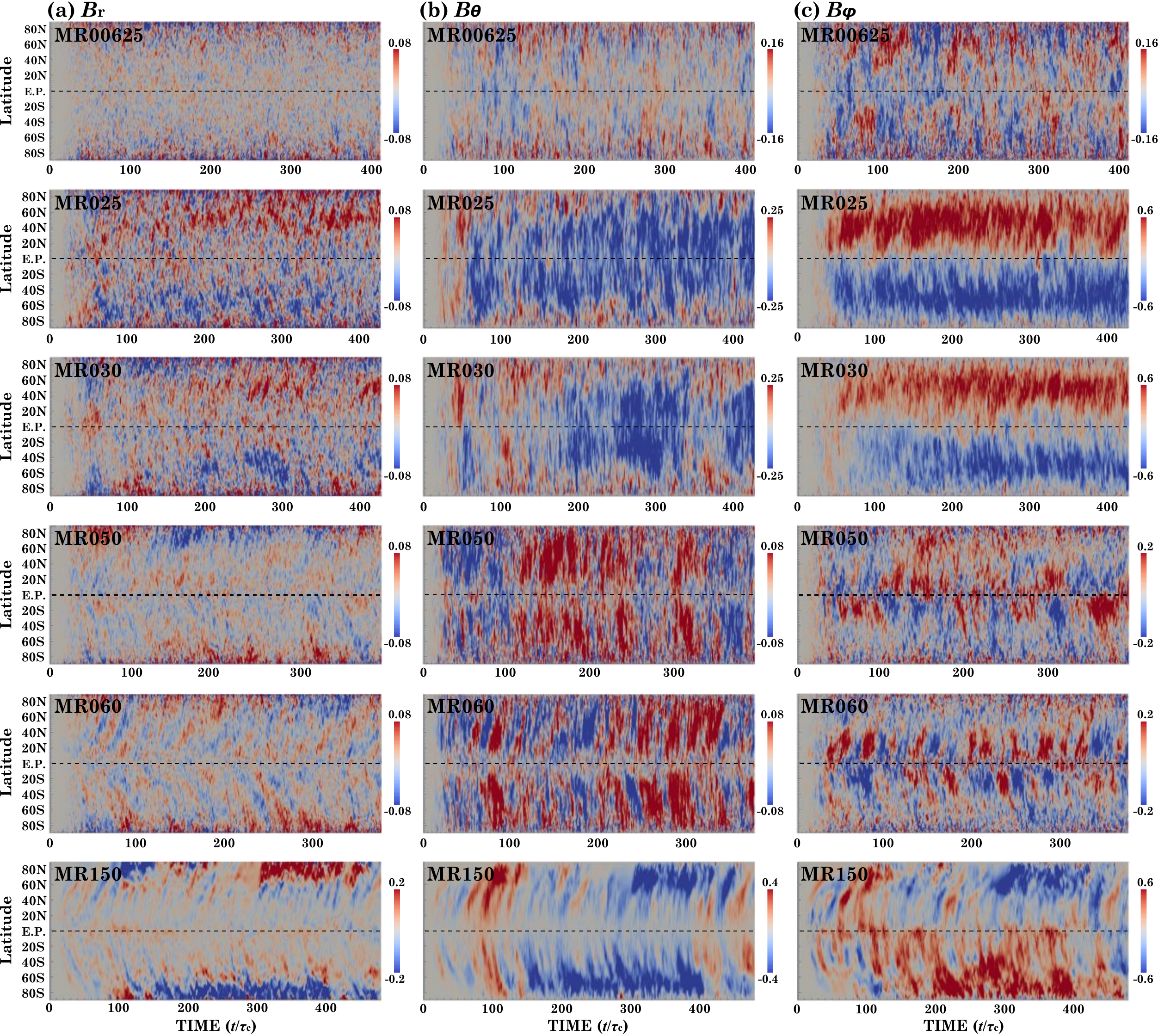}
\caption{Time-latitude diagram of (a) $\langle B_r \rangle_{\phi}$ (left column), (b) $\langle B_\theta \rangle_{\phi}$ (middle column) and 
(c) $\langle B_\phi \rangle_{\phi}$ (right column) at $r = 0.65R$ (mid-stable zone) for the models MR00625, 025, 030, 050, 060, and 150, 
respectively. The time and the amplitude are normalized by $\tau_c$ and $B_{\rm eq}$. The red and blue tones denote the positive and 
negative mean-field strength. Horizontal black dashed-line corresponds to the equator. The top three models have the anti-solar type DR 
and the bottom three models have the solar-type DR.}
\label{fig8}
\end{center}
\end{figure*}
\begin{figure*}[tpb]
\begin{center}
\includegraphics[width=18cm,clip]{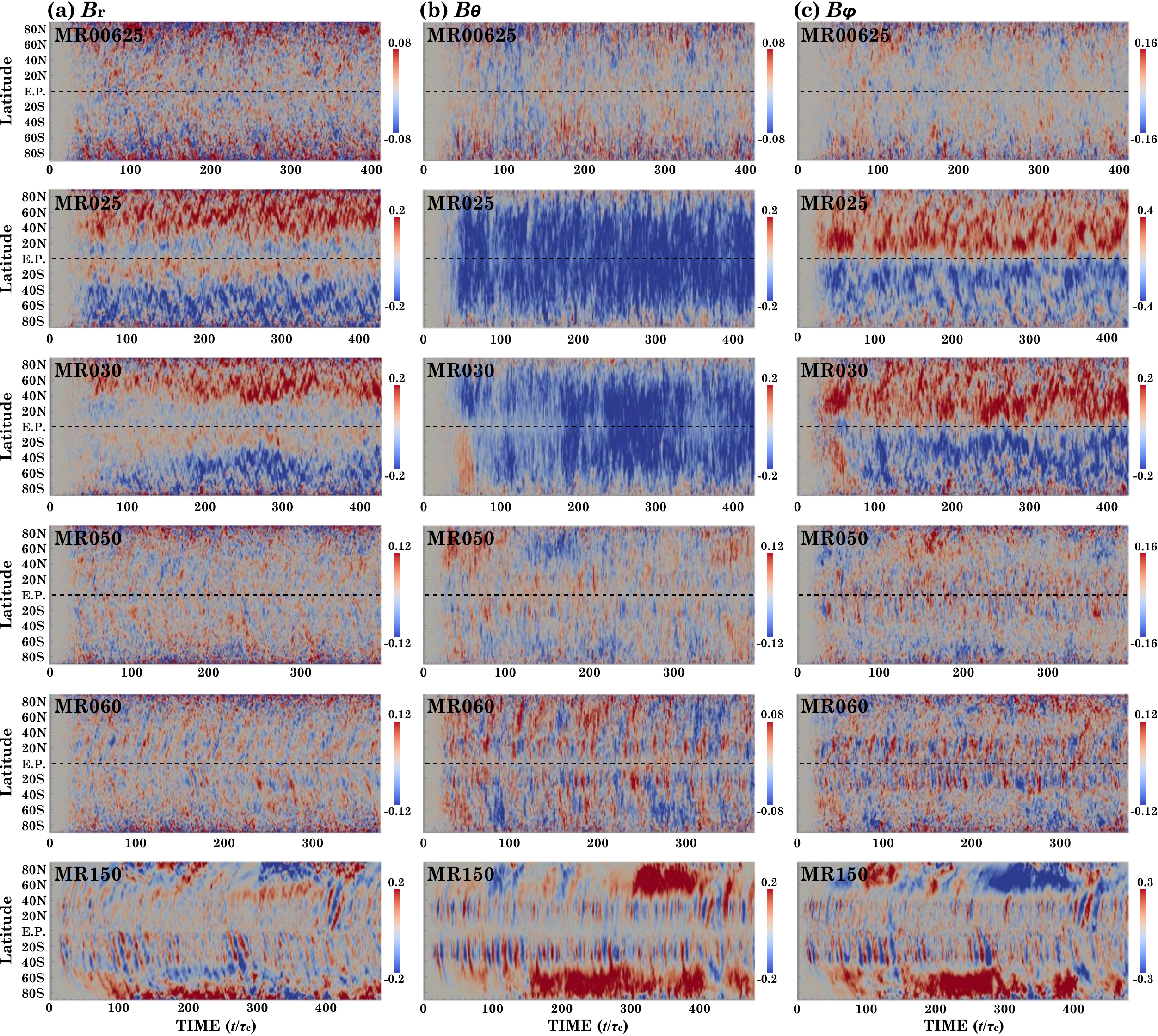}
\caption{Same as Figure 8 but at $r = 0.85R$ (mid-convection zone).}
\label{fig9}
\end{center}
\end{figure*}
\section{Discussion: The Relation Between Magnetic Dynamo Activity and Kinetic Helicity}
\begin{figure}[htpb]
\begin{center}
\includegraphics[width=9cm,clip]{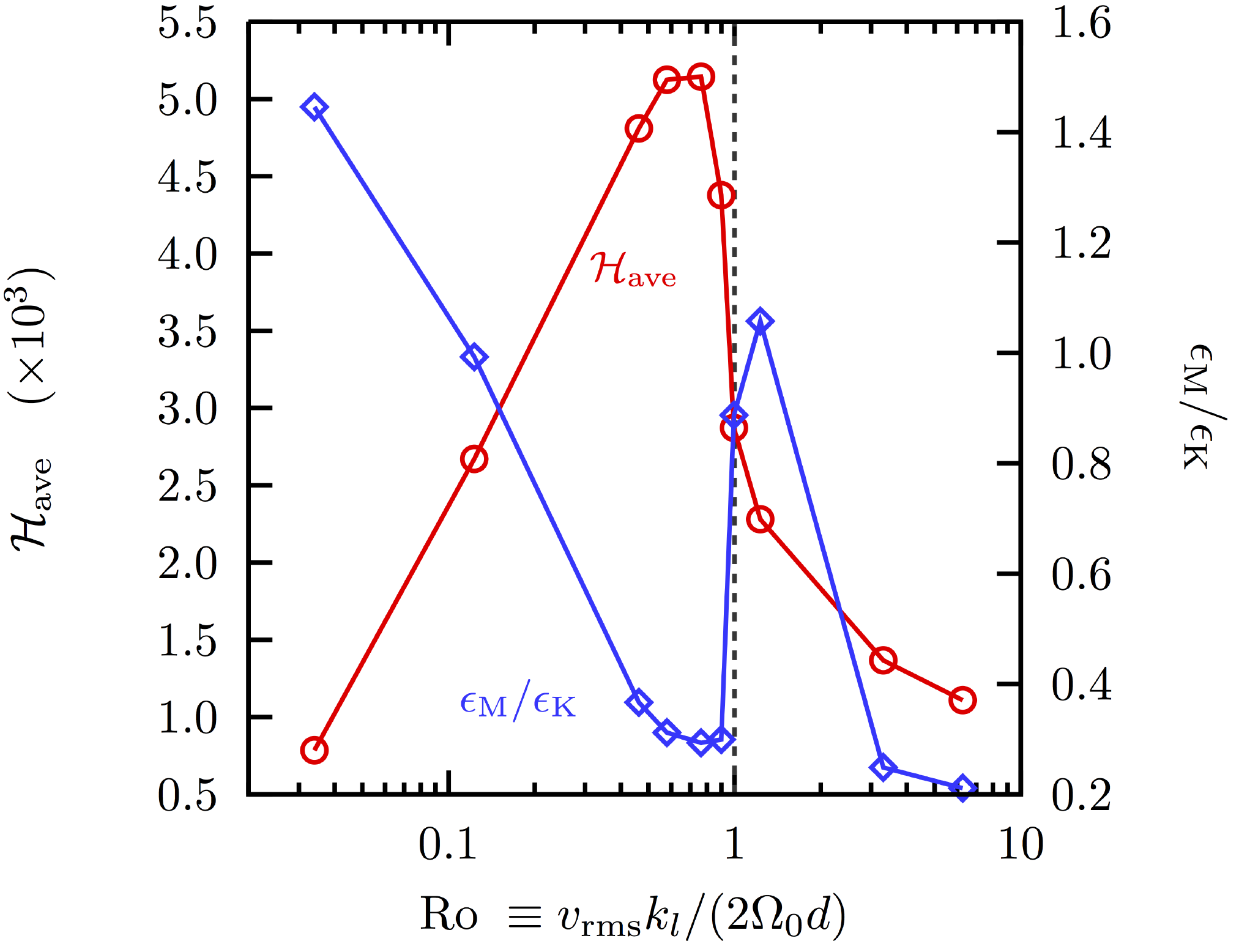}
\caption{Mean helicity defined by $\mathcal{H}_{\rm ave} \equiv (|\mathcal{H}_{\rm N} |+ |\mathcal{H}_{\rm S} |)/2$ (red circles) and 
ratio $\epsilon_M /\epsilon_K$ (blue diamonds) as a function of the Rossby number ($\rm{Ro}$) for the basic MHD runs. }
\label{fig10}
\end{center}
\end{figure}
\begin{figure*}[htpb]
\begin{center}
\includegraphics[width=18cm,clip]{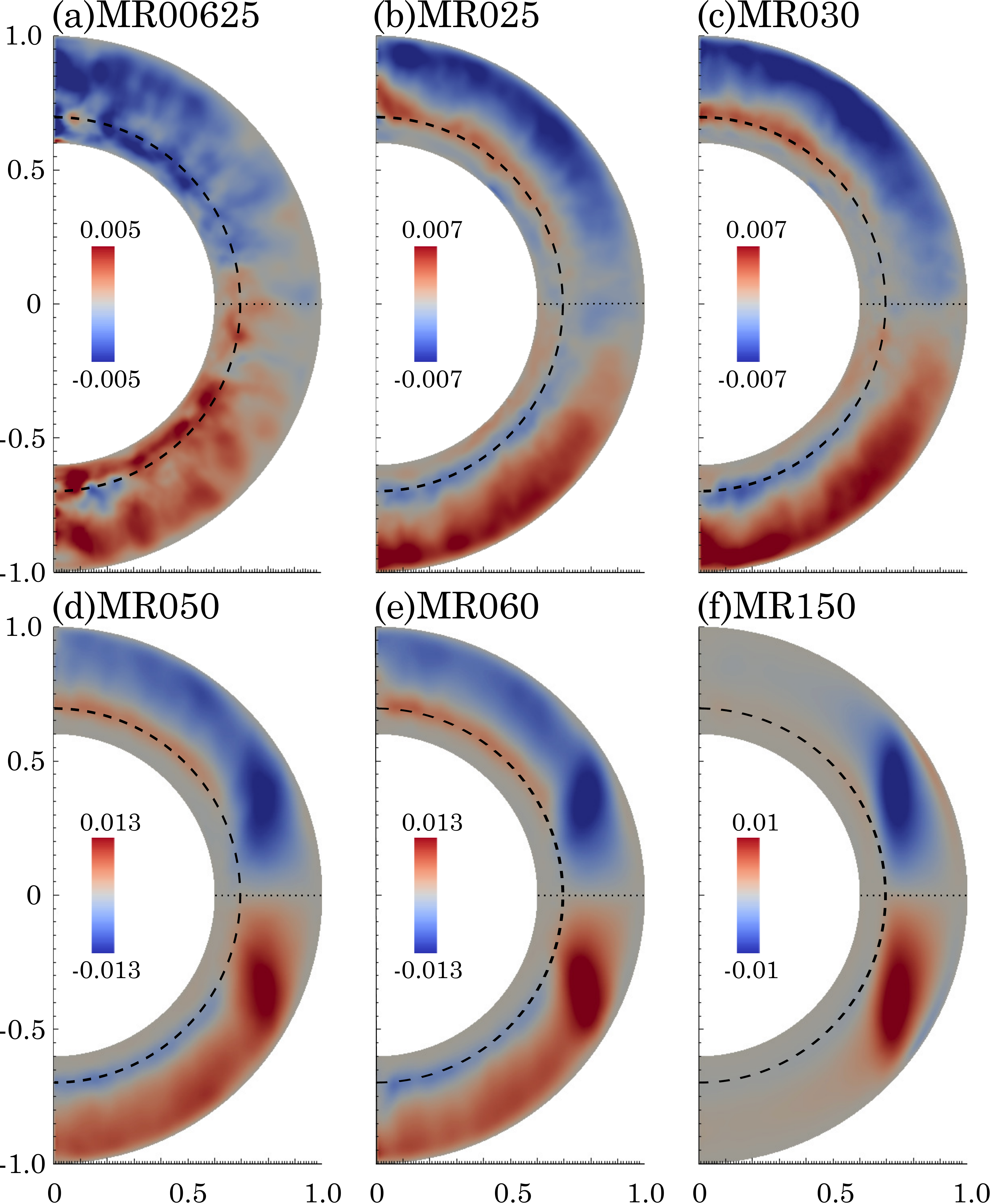}
\caption{Time and azimuthally-averaged kinetic helicity $\overline{\mathcal{H}}(r,\theta)$ for the models 
MR00625, 025, 030, 050, 060, and 150, respectively. The red and blue tones denote the positive and negative helicity. The black 
dotted-line corresponds to the equator. The dashed curve denotes the interface between the convection and radiative zones. The upper 
three models have the anti-solar type DR and the lower three models have the solar-type DR.}
\label{fig11}
\end{center}
\end{figure*}
\begin{figure}[htpb]
\begin{center}
\includegraphics[width=8.5cm,clip]{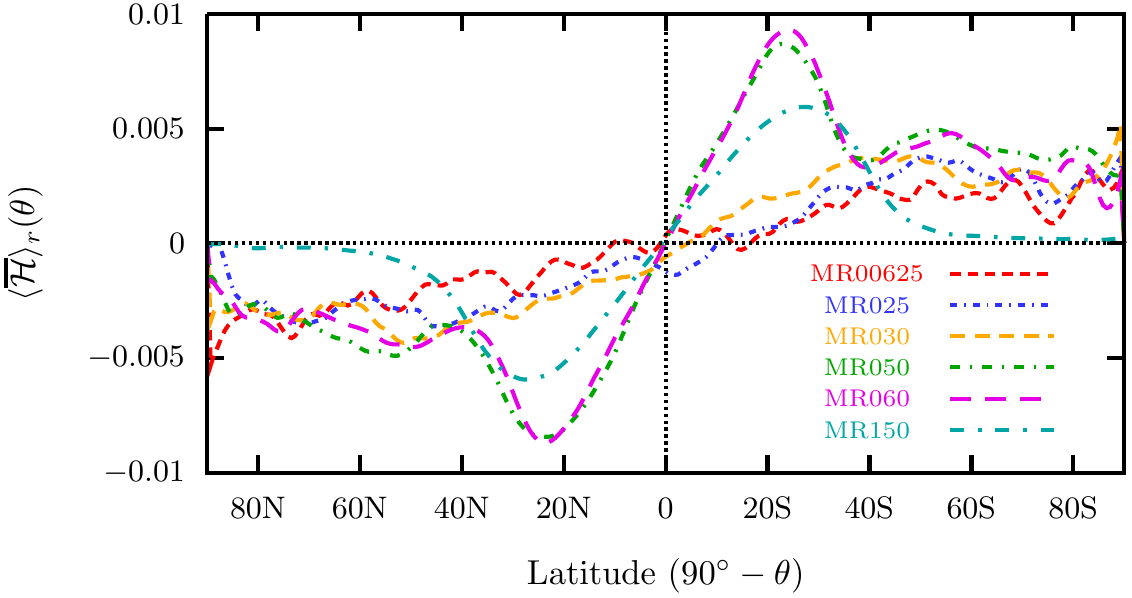}
\caption{Latitudinal distribution of the averaged kinetic helicity, i.e., $\langle \overline{\mathcal{H}} \rangle_r \equiv 
\langle \overline{{\bm v }\cdot (\nabla \times {\bm v})} \rangle_r$. 
The different line types denote the models with different $\Omega_0$.}
\label{fig12}
\end{center}
\end{figure}
The purpose of this paper is to elucidate the effect of the magnetic field on the formation of the rotation profile. Nevertheless, a rich variety 
of the magnetic dynamo activity observed in the MHD models (see \S3.4) makes us interested in their dynamo mechanisms. Here we discuss a 
possible key ingredient for the oscillatory magnetic dynamo with focusing on its relation to the kinetic helicity of the convection flow. 

As an indicator of the magnetic dynamo activity, we study the relative amplitude of the magnetic energy to the kinetic energy which is 
measured by the ratio $\epsilon_{\rm M} /\epsilon_{\rm K}$, where $\epsilon_{\rm M}$ and $\epsilon_{\rm K}$ are mean magnetic and 
kinetic energy densities at the saturated state (see \S3 for the definitions of $\epsilon_{\rm M}$ and $\epsilon_{\rm K}$). In addition, 
the mean kinetic helicity (net helicity), $\mathcal{H}_{\rm ave}$, of each MHD model is also evaluated, where it is defined by 
\begin{equation}
\mathcal{H}_{\rm ave}  =   (|\mathcal{H}_{\rm N}| + |\mathcal{H}_{\rm S}|)/2 \;,
\end{equation}
with 
\begin{equation}
\mathcal{H}_i  \equiv \langle {\bm v }\cdot (\nabla \times {\bm v}) \rangle_i \ \ \ \ \ (i = {\rm N,S}) \;. \nonumber 
\end{equation}
Here the angular brackets with subscripts ``N" and ``S" denote the time- and volume-average in the northern and southern hemispheric 
convection zones at the saturated state. 

Shown in Figure 10 is $\epsilon_{\rm M}/\epsilon_{\rm K}$ (blue line with diamonds [right axis]) and $\mathcal{H}_{\rm ave}$ (red line with 
circles [left axis]) as a function of the Rossby number (${\rm Ro}$) for the basic MHD runs (Set MR). The vertical dashed line denotes 
${\rm Ro} = 1$. The ratio $\epsilon_M /\epsilon_K$ has a bit complicated profile: In the regime of the small or large Rossby number, that is 
${\rm Ro} \ll 1$ or  ${\rm Ro} \gg 1$, $\epsilon_{\rm M}/\epsilon_{\rm K}$ decreases with the increase of ${\rm Ro}$. In contrast, at around 
the transition, the higher ${\rm Ro}$ rather provides the larger $\epsilon_{\rm M}/\epsilon_{\rm K}$. 

The mean kinetic helicity shows a single peak profile with the maximum at around ${\rm Ro} \simeq 1$ where the transition from the solar 
to anti-solar type rotation profiles occurs. In the regime of ${\rm Ro} \lesssim 1$ where the solar-type DR is established, $\epsilon_M /\epsilon_K$ 
anti-correlates with $\mathcal{H}_{\rm ave}$. By contrast, the ratio decreases with the kinetic helicity in the anti-solar type rotation regime of 
${\rm Ro} \gtrsim 1$. Since the temporal property of the large-scale magnetic field also changes from the oscillatory to the stationary across 
${\rm Ro} \simeq 1$ as was described in \S3.4, this suggests that the dynamo mechanism would be associated in some way with the kinetic 
helicity in the convection zone. 


The ${\rm Ro}$-dependence of the net kinetic helicity, which shows a peak at moderate rotation rate, is deeply related to the 
formation of the convective column. Since the angular velocity is lower in the regime of the anti-solar type DR or ${\rm Ro} \gg 1$, the up-down 
asymmetry of the convective motion is lower, leading to less preference for one sign of the kinetic helicity. On the other hand, in the regime of 
the solar-type DR or ${\rm Ro} \ll 1$, the convection has a tendency to align in columnar rolls parallel to the rotation axis (see \S3.1). In the convection 
columns, the velocity is mostly perpendicular to the rotation axis while the vorticity is mostly parallel to the rotation axis, resulting in a small 
net kinetic helicity \citep[e.g.,][]{knobloch+81,browning08}. It was also clearly demonstrated by \citet{sprague+06} with asymptotic numerical 
analysis of unstratified turbulence that the largest net kinetic helicity is established at moderate rotation rate and decreases as rotation becomes 
even more rapid \citep[see also,][]{kapyla+09a}. 

The anti-correlation between ${\rm Ro}$ and $\epsilon_M /\epsilon_K$ in the solar-type DR regime would be consistent with the earlier 
studies \citep[e.g.,][]{brown+11,kapyla+13}. In addition, the correlation between ${\rm Ro}$ and $\epsilon_M /\epsilon_K$ around the transition is also 
compatible with the numerical result of \citet{karak+14} (see their Table 2). However, 
the role of the kinetic helicity in the magnetic dynamo mechanism and its hidden connection to the observed relationship between stellar 
rotation and magnetic activity \citep[e.g., see][for reviews]{gudel07,reiners12} remain to be fully elucidated although there are a great deal of work 
exploring the linkage between them \citep[e.g.,][]{knobloch+81,jennings+91,baliunas+96,brandenburg+98,brandenburg+12}.

To get a better grasp of the relation of the kinetic helicity and the magnetic dynamo mechanism, in Figure 11, we show the distribution of the time 
and azimuthally-averaged kinetic helicity, $\overline{\mathcal{H}}(r,\theta) \equiv \overline{{\bm v }\cdot (\nabla \times {\bm v})}$, for the models 
MR00625, 025, 030, 050, 060, and 150, respectively. The red and blue tones denote the positive and negative helicity. The horizontal black 
dotted-line corresponds to the equator. The dashed curve denotes the interface between the convection and radiative zones. The models listed 
in the upper three panels have the anti-solar type DR (MR00625, 025 and 030) and the lower three panels correspond to the models with the solar-type DR 
(MR050, 060 and 150). To clearly demonstrate the latitudinal distribution of the kinetic helicity, the radial average 
of the time- and azimuthally-averaged kinetic helicity, i.e., $\langle \overline{\mathcal{H}} \rangle_r \equiv 
\langle \overline{{\bm v }\cdot (\nabla \times {\bm v})} \rangle_r$, is also shown in Figure 12, where 
angular brackets with subscript $r$ denote the radial average. The different line-types denote the models with different $\Omega_0$.

It would be trivial that the kinetic helicity has an antisymmetric profile with respect to the equator in the model with a sufficiently fast rotation. 
In the northern (southern) hemisphere, the negative (positive) helicity becomes predominant. This is because the downflow with the negative 
(positive) helicity is on average stronger than the upflow with the positive (negative) helicity in the northern (southern) hemisphere in the density 
stratified system \citep[e.g.,][]{spruit+90,miesch05}. The weaker equatorial antisymmetry of the kinetic helicity profile in the model with the smallest 
rotation (MR00625) would be a result of a weaker Coriolis force which provides less impact on the convective motion. 
The peak amplitude of the kinetic helicity is little dependent on the rotation rate (within factor 2). However, we can find from Figure 12 that the latitude 
for the peak of the kinetic helicity decreases as the rotation rate increases: While the higher latitudinal region has a larger helicity in the models with the 
anti-solar type DR (MR00625, 025, and 030), the lower latitudinal region has a larger helicity in the models with the solar-type DR (MR050, 060, and 150). 

It is the most remarkable that both the hemispheric and inter-hemispheric latitudinal gradients of the kinetic helicity become larger around the equator 
with the increase of the rotation rate. In the models with the solar-type DR and the oscillatory large-scale magnetic field, the kinetic helicity concentrates 
on around the equatorial plane and has strong hemispheric and inter-hemispheric latitudinal gradients there. In contrast, the models with the anti-solar type 
DR and the stationary large-scale magnetic field do not show the significant latitudinal gradient of the kinetic helicity around the equatorial region. 
Since, when comparing Figures 9 and 11, the latitude providing the stronger large-scale magnetic field overlaps with that with the larger kinetic helicity, 
it is expected that the latitudinal distribution of the kinetic helicity is a key to distinguish the dynamo mechanisms. The larger the latitudinal gradient of the 
kinetic helicity around the equator becomes, the more the oscillatory large-scale magnetic field would be developed. 

Recently, \citet{mitra+10} reported an intriguing finding in their dynamo simulation by a forced helical turbulence in a wedge-shaped spherical shell that the 
oscillatory large-scale magnetic field with equatorial migration can be organized, even without the $\Omega$-effect, essentially due to an inter-hemispheric 
gradient of the kinetic helicity. They argued that the $\alpha^2$-dynamo mode excited by the helical turbulence with the equatorial antisymmetry is responsible 
for the oscillatory property. Furthermore, a connection between $\alpha^2$-dynamo mode and solar magnetism was discussed in some recent results of 
convective MHD dynamos simulations \citep[e.g.,][]{simard+13,masada+14a,masada+14b}. 

When considering the latitudinal gradient of the kinetic helicity around the equator as an important ingredient of the oscillatory property of the dynamo even 
in our simulations, we can speculate the reason why there exists a difference of the dynamo property depending on the regime of the rotation profile. Since 
the model with the smaller rotation rate and thus the anti-solar type DR does not have a sufficient latitudinal gradient of the kinetic helicity around the equator 
to excite the oscillatory $\alpha^2$-dynamo mode, the $\Omega$-effect solely works to build up the large-scale magnetic field and thus yields a stationary 
solution. In contrast, the model with the faster rotation and thus the solar-type DR has a sufficient concentration and strong latitudinal gradient of the kinetic 
helicity around the equator, yielding the oscillatory large-scale magnetic component sustained by the $\alpha^2$-effect in addition to the $\Omega$-effect. 

The detailed analysis with mode expansion by Legendre polynomials on kinetic helicity and magnetic field will be presented in a subsequent paper 
in which we focus on the magnetic dynamo activity and its dependences on the rotation rate and stratification level. Although a quantitative description of 
the dynamo mechanism is beyond the scope of this paper, we can speculate from our simulation results that accurate numerical modeling of the kinetic helicity 
profile in the actual solar interior will offer a way to unveil the mystery of the solar magnetism. 
\section{Summary}
In this paper, the effect of the magnetic field on the mean DR profile established in the convection zone was systematically studied by 
comparing the MHD and HD models of the rotating spherical shell convection in the broad range of the initial rotation rate. 
A fully compressible Yin-Yang MHD code which was developed in MYK13 and a stellar model consisting of the convection zone and the 
radiative zone were used for the simulation. The critical parameter which controls the transition between the anti-solar type and solar-type 
DR profiles was explored with focusing on the ``Rossby number (${\rm Ro}$)" and the ``convective Rossby number (${\rm Ro}_{\rm conv}$)". 
In addition, the bistability of the rotation profile, which has only recently been discovered by \citet{gastine+14}, was revisited under the 
influence of the magnetic field. Our main findings are summarized as follows. 

1. The critical value of the ${\rm Ro}_{\rm conv}$ for the transition was higher in the MHD model than the HD model. The 
transition was more gradual in the model under the influence of the magnetic field. Comparing two models at the same 
${\rm Ro}_{\rm conv}$, the magnitude of the DR parameter ($\alpha_e$) was smaller in the MHD model than in the HD model. 
The magnetic field had a crucial role in suppressing the DR and facilitated to produce the solar-type rotation profile. Since the Lorentz force 
of the dynamo-generated magnetic field reduced the convective velocity, the MHD model became more rotationally-constrained than the 
HD model. This would be the reason why the magnetic field facilitated to produce the solar-type DR. 

2. The transition from the solar to anti-solar type rotation profiles occurred at ${\rm Ro} \simeq 1.0$ both in the MHD and HD models. 
Not only the critical value, the sharpness of the transition was also similar between the two models in the view of ${\rm Ro}$-dependence. 
The solar-type DR, accompanied by the columnar convection and multi-cell meridional circulation pattern, was established in the regime 
${\rm Ro} \lesssim 1$. The anti-solar type DR, characterized by the cellular convections and the single-cell meridional circulation, developed 
in the regime of ${\rm Ro} \gtrsim 1$ regardless of the presence of the magnetic field. The transition between the two rotation profiles was 
controlled by the ${\rm Ro}$ rather than by the ${\rm Ro}_{\rm conv}$. 

3. From the point of view of the ${\rm Ro}_{\rm conv}$-dependence, the rotation profile showed, as was observed in 
earlier studies, the bistability near the transition in the HD model while it disapeared when allowing the growth of the 
dynamo-generated magnetic fields except for the model with taking an anti-solar type solution as the initial condition. It was found 
that the transition between the anti-solar and solar-type DR profiles could be unified in the view of ${\rm Ro}$-dependence 
independently of the hysteresis. The transition between the two rotation regimes was again better captured by the ${\rm Ro}$ rather 
than by the ${\rm Ro}_{\rm conv}$.

4. The influence of the magnetic field or the history on the convective velocity would be responsible for the dependence 
of the rotation profile on it. Such effects were not reflected in the ${\rm Ro}_{\rm conv}$ by definition. In contrast, it could be definitely 
captured by the ${\rm Ro}$. The transition between the solar and anti-solar type rotation profiles was thus unified in the view 
of the ${\rm Ro}$-dependence. 

5. A rich variety of the magnetic dynamo activity could be observed in the MHD models. In the regime of the anti-solar type DR 
(${\rm Ro} \gtrsim 1$), the stationary large-scale magnetic field with a dipole symmetry was built up both in the stable and convection 
layers except for the model with the smallest rotation rate. In contrast, the oscillatory large-scale magnetic component was organized 
in the model with the solar-type DR profile (${\rm Ro} \lesssim 1$). The latitudinal profile of the kinetic helicity would be a key to 
distinguish the dynamo mechanisms between the two rotation regimes. The oscillatory large-scale magnetic field could be organized 
when the strong latitudinal gradient of the kinetic helicity existed around the equator. 

Our parametric study was conducted by the simulation model with the same value of the Rayleigh number. The supercriticality level $\delta$ 
($\equiv [{\rm Ra} - {\rm Ra}_c]/{\rm Ra}_c $), which is an essential measure of the turbulence level, was thus not constant and varied with 
the rotation rate as shown in Table 1 (see the third column). In order to compare the models with the supercriticality level fixed, we should 
change the Ekman number or Prandtl number, or density stratification, providing a possible difference in the resultant flow pattern. 

Gastine et al. (2013,2014) indicated that, while the transition between the solar and anti-solar type rotation DR profiles would take place 
at almost the same ${\rm Ro}_{\rm conv}$ independently of the density stratification and Ekman number, the lower Ekman number or the 
stronger density stratification would produce the steeper transition. It was also suggested that the higher Prandtl number would shifts the 
critical ${\rm Ro}_{\rm conv}$ to the higher value at least in the high Ekman number regime. Not only on the flow properties, these parameters 
would also have an influence on the magnetic dynamo activities \citep[c.f.,][]{kapyla+11,kapyla+14,karak+14}. To verify the universality of the 
critical Rossby number for the transition obtained in this study, further parameter study is indispensable and is planned as our future work. 

The recent development of the astroseismology opens up the way to study the large-scale internal flows in the solar-type main sequence 
stars with different age and thus different rotation rate \citep[e.g.,][]{chaplin+10}. Computer simulation in tandem with the advanced 
observation will help deepening the understanding of the stellar interior dynamics and stellar dynamo activities in the astroseismology era.
\acknowledgments
We thank the anonymous referees for constructive comments. Numerical computations were carried out on Cray XC30 at Center for 
Computational Astrophysics, National Astronomical Observatory of Japan. 
This work was supported by JSPS KAKENHI grant Nos. 24740125 and 20260052.  

\clearpage
\end{document}